\documentclass[12pt]{iopart}
\usepackage{iopams}

\expandafter\let\csname equation*\endcsname\relax

\expandafter\let\csname endequation*\endcsname\relax

\usepackage{graphicx}
\usepackage{dcolumn}
\usepackage{bm}

\usepackage{amsmath}
\usepackage{amsthm,amsfonts, amssymb}

\usepackage[usenames, dvipsnames]{xcolor}
\usepackage{array}
\usepackage{makecell}

\usepackage{verbatim}
\usepackage{dsfont}

\makeatletter
\newcommand*{\transpose}{%
	{\mathpalette\@transpose{}}%
}
\newcommand*{\@transpose}[2]{%
	\raisebox{\depth}{$\m@th#1\intercal$}%
}
\makeatother

\newcommand{\bra}[1]{\langle #1 |}
\newcommand{\ket}[1]{| #1 \rangle}
\newcommand{\Pproj}{\mathcal{P}}
\newcommand{\Qproj}{\mathcal{Q}}
\newcommand{\Trb}[1]{\textnormal{Tr}_B\{#1\}}

\usepackage{mdframed}
\newcommand{\tboxit}[2]{
\vspace{2mm}
\begin{mdframed}[
        linecolor=black,linewidth=0.75pt,%
        frametitlerule=true,%
        frametitlerulecolor=black,
        frametitlerulewidth=0.5pt, innertopmargin=\topskip,
        frametitle={\begin{center}\scshape \mdseries #1\end{center}},
        outerlinewidth=1pt,
        innerbottommargin = 15pt
    ]
    #2 
\end{mdframed}
\vspace{2mm             }
}

\usepackage{tikz}

\usetikzlibrary{calc}

\usetikzlibrary{positioning}
\tikzset{node distance=2cm, auto}
\begin{document}
	\title[Efficient energy resolved quantum master equation]{Efficient energy resolved quantum master equation for transport calculations in large strongly correlated systems}
	
	\author{Gerhard Dorn$^1$, Enrico Arrigoni$^1$ and
		Wolfgang von der Linden$^1$}
	
	\address{$^1$ Institute of Theoretical and Computational Physics, Graz University of Technology, NAWI Graz, 8010 Graz, Austria}
	\ead{quantum.quentin@mailbox.org}
	
	\begin{abstract}
	We introduce a systematic approximation for an efficient evaluation of Born--Markov master equations for steady state transport studies in open quantum systems out of equilibrium: the energy resolved master equation approach. The master equation is formulated in the eigenbasis of the open quantum system and build successively by including eigenstates with increasing grandcanonical energies. In order to quantify convergence of the approximate scheme we introduce quality factors to check preservation of trace, positivity and hermiticity. 
	
	Furthermore, we discuss different types of master equations that go beyond the commonly used secular approximation in order to resolve coherences between quasi--degenerate states. For the discussion of complete positivity we introduce a canonical Redfield-Bloch master equation and compare it to a previously derived master equations in Lindblad form with and without using the secular approximation. The approximate scheme is benchmarked for a six orbital quantum system which shows destructive quantum interference under the application of a bias voltage. The energy resolved master equation approach presented here makes quantum transport calculations in many--body quantum systems numerically accessible also beyond six orbitals with a full Hilbert space of the order of $\sim 10^6$.
	\end{abstract}
\noindent{Keywords}:	quantum master equation, efficient evaluation, destructive interference, complete positivity
	\submitto{ Quantum Sci. Technol.}
	\maketitle

\date{\today}



\section{\label{sec:level1}Introduction}
In the breakdown of Moore's law when reaching the quantum limit, the study of quantum effects on electronic transport is becoming increasingly important. The quantum nature of electrons is not only a source of error for logical circuits, but also opens up a wide range of new applications, provided that one is able to correctly predict, understand and enhance these quantum effects, which mostly arise at small scales and low temperatures. 
One of the most successful methods to derive an appropriate description of the electronic landscape for such systems in that parameter regime is the non--equilibrium density functional theory (DFT) in combination with non--equilibrium Green function (NEGF) techniques \cite{papior2017improvements, novaes2006density}. However, its mean-field nature and the diagrammatic difficulties in the calculation of the correction term (self-energy) have major disadvantages when investigating transport effects arising in a system of strongly correlated electrons.

In the case of a small strongly correlated system weakly coupled to a large non--interacting environment one can use the concept of so--called open quantum systems. In order to determine the actual state of the quantum system an effective differential equation (master equation) and effective time evolution (dynamical map) have to be derived. The theoretical study of completely positive, trace and hermiticity preserving linear maps of reduced density matrices lead to the formulation of the most generic form of such a master equation, the famous Gorini--Kossakowski-Gorini-Sudarshan-Lindblad equation~\cite{gorini1976completely,Lindblad1976} [see also~(\ref{equ:lindblad_equation})], which consists of an effective Hamiltonian describing the altered system dynamics and a dissipative term encoding the influence of the environment. 

Since most setups cannot be mapped onto an exact master equation there are two common methods to derive an approximate one~\cite{breuer2009measure, whitney2008staying}. One is a phenomenological approach, which starts from the Lindblad equation and tries to derive the unknown parameters (e.g. estimating the decoherence rates in optical cavities). The other is a microscopic derivation where one starts from a model Hamiltonian (e.g. derived from DFT ab--initio calculations in molecular electronics) and tries to find a Lindblad--like master equation by applying perturbation theory and further approximations.


Because the latter approach may face problems concerning the perturbative character of the derivation and in some cases it does not lead to a Lindblad equation, so that complete positivity is violated, a lot of research has been done to tackle those problems. 
Recent developments in the microscopic derivation of quantum master equations such as hierarchical quantum master equation~\cite{wenderoth2016sharp,schinabeck2016hierarchical} or auxiliary master equation~\cite{DordaNussEtAl2014,sorantin2019auxiliary} try to avoid perturbative methods to enhance the applicability to stronger values of the system--bath couplings at the cost of an expensive calculation limiting the method to rather small quantum systems (single or double impurities or qubits). On the other hand, several works focus on the positivity issue in the weak coupling regime discussing time-local master equations~\cite{whitney2008staying, schultz2009quantum,  jeske2015bloch, eastham2016bath, dominy2016beyond}. 

The focus of this work is to show how time--local master equations 
can be efficiently evaluated to describe steady state dynamics for rather large systems. 
The numerical challenge for treating those larger interacting quantum many--body systems with master equations lies in the fast growth of the size of the Hilbert space. Both the size $n$ of the full density matrix of a system with electron--electron interactions ($n = 4^l$) as well as the number of equations $\mathcal{N}$ in the master equation ($\mathcal{N} = 4^{2l}$) scale exponentially with increasing number of systems sites (orbitals) $l$. 
To climb this exponential wall  we here introduce a scheme to suitably truncate such large Hilbert spaces, the so-called energy resolved master equation approach (ERMEA). This approach provides a systematic approximation valid for the most common master equations in the weak--coupling regime.

ERMEA makes master equation techniques especially applicable to transport simulations of larger ab--initio derived strongly correlated quantum systems that are attached to baths with arbitrary densities of states.

In order to validate the approximation and convergence of the master equation (also referred to as generator of a dynamical map) we introduce quality factors, measuring the preservation of trace, hermiticity and (complete) positivity  for the master equation induced map and the convergence of the steady state.

\paragraph{Positivity issue and coherence phenomena:} The inclusion of coherences between different eigenenergies has in many derivations of master equations the price of loosing complete positivity (as the secular approximation is not applicable) which was also discussed in~\cite{whitney2008staying, jeske2015bloch, dominy2016beyond}. 
After summarizing different options of master equations for describing those coherences we introduce a new type of Redfield--Bloch type master equation that helps to quantify this violation of complete positivity and possibly the violation of Markovianity in an easy calculable way.

The article is structured as follows:

In Sec.~\ref{sec:dynamical_map} we give a short introduction to open quantum systems and dynamical maps, that clarifies for which parameter regimes the suggested evaluation method (ERMEA) is suited.

The energy resolved master equation evaluation approach and all relevant quality factors are introduced and discussed  in Sec.~\ref{sec:ERMEA}.

Sec.~\ref{sec:master_equations} is dedicated to different types of Born--Markov master equations that can be used in combination with ERMEA. This section contributes to the discussion of positivity in master equation derivations.

For testing ERMEA in combinaton with various master equations and in order to emphasize the need to include possible coherences, we discuss an interesting setup of destructive quantum interference - current characteristics of a benzene molecule connected in the meta configuration~\cite{begemann2009quantum} - in Sec.~\ref{sec:num_res_ERMEA}.

\section{Dynamical map - the theory in a nutshell}\label{sec:dynamical_map}
The following section gives a short introduction to the basics of a dynamical map and introduces most terms used in this context.
For a comprehensive introduction to the field of quantum dynamical maps, uniformly continuous semigroups and quantum master equations we refer to the detailed review of Rivas and Huelga \cite{rivas2012open}. 

\subsection{General concept of an open quantum system and a dynamical map}
In general an \textbf{open quantum system} consists of a quantum system of interest ($S$), that is connected via a (weak) coupling ($I$) to a bath or environment ($B$). The total system will be described by a Hamiltonian consisting of the three parts: system, bath and coupling (perturbation) Hamiltonian:
$$ H = H_S + H_B + H_I,\quad H_0 := H_S + H_B. $$ 
System and bath Hamiltonians together form the decoupled Hamiltonian $H_0$. Convenient choices for the baths are non--interacting leads in asymptotically thermal equilibrium. These could be realized as semi--infinite tight binding chains or as flat bands in the wide band limit. The non--equilibrium situation of applying a bias voltage can be realized by shifting the chemical potentials in the baths. 

\paragraph{Differential picture:} The fundamental question is how to derive an effective description of the open quantum system of interest $S$ under the influence of the environment $B$. 
A full characterization of $S$ is provided by the reduced density operator $\sigma(t)$ which can be obtained from the full density operator $\rho(t)$ by tracing out the bath degrees of freedom $ \sigma(t)= \Trb{\rho(t)}$. 

In order to describe the dynamics of such a system $S$ under the influence of the baths an effective differential equation, in this context referred to as \textbf{quantum master equation}, of the following time--local form is searched: 

\begin{align}
    \frac{\rmd\sigma(t)}{\rmd t} &= \mathcal{K}(\rho_B(t), t)\{\sigma(t)\}, \label{equ:superoperator}\\
    \sigma(t_0) &= \sigma_0.
\end{align}
The central part of the master equation is the \textbf{generator} $\mathcal{K}(\rho_B(t),t)\colon \mathcal{H} \rightarrow \mathcal{H}$ that maps from the Hilbert space $\mathcal{H} = \mathbb{C}^{n\times n}$ of reduced density operators onto itself and is thus also called superoperator. The generator incorporates the influence of the fixed bath $\rho_B$ and is in general a time-dependent superoperator that acts non-linearly (not time-locally) on $\sigma(t)$. 

In principle such a time--local (or time--convolutionless) differential equation for some part of a quantum system does not always exist or may not be unique since the information of already present \textbf{correlations} between system and environment is missing. One can show that for the special case of having no such correlations at some initial time $t_0$ where the quantum system can be written as a product state $\rho(t_0) = \sigma(t_0) \otimes \rho_B(t_0)$, it is possible to formulate a time--local quantum master equation valid at $t_0$ using a Kraus operator decomposition~\cite{kraus1983states}.

The emergence of correlations which prohibit a straightforward formulation of master equation are related to memory effects which are linked to non--Markovian behaviour. A strategy to take those memory effects into account is deriving a non--time--local (or non--convolutionless) master equation starting from a time with zero correlations. One should mention that in many cases a time--local master equation can be derived also in a non--Markovian setup~\cite{hall2014canonical}.

\paragraph{Picture of time evolution:} The solution $\mathcal{U}(t, t_0, \rho_B(t_0), \sigma(t_0))$ of the quantum master equation is called \textbf{dynamical map}. It incorporates the influence of the bath and of memory effects and provides an effective time evolution for $S$:
\begin{align}
    \sigma(t_1) = \mathcal{U}\left(t_1,t_0,\rho_B(t_0), \sigma(t_0)\right) \,\sigma(t_0),
\end{align} which is related to the full unitary time evolution $U$ via partial traces over the bath degrees of freedom
$$ \sigma(t_1) = \Trb{U(t_1, t_0) \rho(t_0)}. $$
For a shorter notation we omit from now on the dependence on the bath in the dynamical map: $\mathcal{U}(t_1, t_0, \sigma(t_0))$.
\subsection{Necessary conditions for a dynamical map and reasonable simplifications}
\begin{figure}[ht!] 
 \begin{center}
	\includegraphics[width=0.8\textwidth]{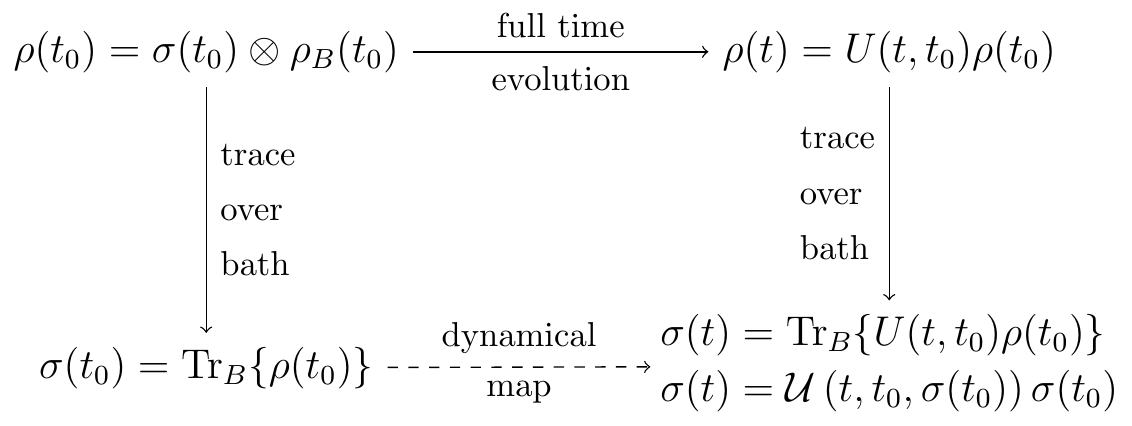}
 \caption{Comparison of a trace--induced dynamical map to the full time evolution for an initial product state.}
 \label{fig:dynamical_map}
 \end{center} 
\end{figure}

 Figure~\ref{fig:dynamical_map} compares the dynamical map induced by tracing out the baths into a full unitary time evolution $U$ for an initial product state.
In general such an effective time evolution (dynamical map) should fulfill the following conditions:
\begin{itemize}
    \item[Con~1.] Preserve the trace of the reduced density operator $\sigma(t)$,
    \item[Con~2.] Preserve the hermiticity of $\sigma(t)$,
    \item[Con~3.] Preserve the positivity of $\sigma(t)$,
    \item[Con~4.] Divisibility: $$\mathcal{U}(t_2,t_0, \sigma(t_0)) = \mathcal{U}(t_2,t_1, \sigma(t_1)) \circ \mathcal{U}(t_1,t_0,\sigma(t_0)).$$
\end{itemize}
In the following we will discuss which assumptions simplify the formulation of a dynamical map with the listed properties.
\paragraph{Independence / linear generator:} It is desirable to have a dynamical map $\mathcal{U}(t_1,t_0)$ which is independent of the initial density matrix and basically works for all initial values $\sigma(t_0)$. In the differential picture this amounts of having a generator $\mathcal{K}(t)$ acting linearly on $\sigma$. It can be shown~\cite[Th.~4.3]{rivas2012open}, that this holds true under the following condition, namely, that there is a time $t_0$ at which $\rho(t_0)$ can be written as a tensor product $\sigma(t_0) \otimes \rho_B(t_0)$ between system and bath (no quantum or classical correlations at the beginning) where the bath $\rho_B(t_0)$ is the same for any $\sigma(t_0)$.

This independence from the initial value and applicability of the dynamical map on all initial states motivates the definition of a so called \textbf{universal dynamical map} (UDM), which requires besides Con.~1 and Con.~2 in addition \textbf{complete positivity}, which means that for any extra system of arbitrary dimensionality $d$ the evolution $\mathcal{U}(t_1,t_0)\otimes \mathds{1}_d$ also  preserves positivity~\cite{nielsen2002quantum}, in other words preservation of positivity is independent of the possible entanglement with an attached environment. As a counterexample, one can mention the transposition operator, which while being positive violates complete positivity.


\paragraph{Divisibility: } It is important to note that correlations building up between bath and system are an obstacle for the desired divisibility (Con.~4) with UDMs. Assuming a tensor product as initial state at $t_0$, a UDM can be formulated between $t_0 \rightarrow t_1$ and $t_0 \rightarrow t_2$. Because of the possible correlations it is per se not possible to have a UDM from $t_1 \rightarrow t_2$. 


In the weak coupling regime the so-called \textbf{Born approximation} is applicable in an expansion term (see \ref{sec:app_sub_born_approx}) of the master equation, which enforces divisibility by arguing that arising correlations between bath and system can be neglected ($\rho(t) \approx \sigma(t) \otimes \rho_B$) under certain conditions. The Born approximation (assumption of state factorization for all times in the integral term) can also be understood as an effective model when deriving the master equation using projection operator techniques~\cite{rivas2010markovian}.

\paragraph{Steady state, time independence:} In this work we are interested in the steady state properties of the system under investigation thus we assume not only a linear but also a time-independent superoperator $\mathcal{K}$ in~(\ref{equ:superoperator}). As a consequence the  effective time evolution operator will only depend on the time difference $$\sigma(t) = \mathcal{U}(t-t_0) \sigma (t_0) = \rme^{-\rmi \mathcal{K}(t-t_0)} \sigma(t_0).$$ 


The eigenvector of the linear superoperator $\mathcal{K}$ with eigenvalue zero is the steady state of the system. This will be the object of investigation in this work.
\subsection{General form of a master equation generating a universal dynamical map}
Having defined the general requirements of a UDM we now examine their implications on the form of its generator, the master equation.
Demanding Con.~1--3 and complete positivity to hold, leads consequently to the most general time independent form of a universal dynamical map, the so-called Gorini--Kossakowski-Sudarshan-Lindblad equation~\cite{gorini1976completely,Lindblad1976} equation:
\tboxit{Lindblad form}{
\begin{align}
    \dot{\sigma}(t) =& -\frac{\rmi}{\hbar} [H_S + H_{\textnormal{LS}},\sigma(t)] +\sum_{ij} \Gamma_{ij} \left(A_i \sigma(t) A_j^\dag -\frac{1}{2} \{A_j^\dag A_i, \sigma(t)\}\right). \label{equ:lindblad_equation}
\end{align}
with the orthonormal (Hilbert-Schmidt norm) basis of trace-less 
operators $A_i$ (see for example \ref{sec:canonical_Redfield_Bloch}) and the indices $i\in[1,\dots n^2-1]$ with $n$ the dimension of the Hilbert space of the system (number of eigenstates). The matrix $\Gamma$  incorporates the dissipative influence of the bath, whereas the so--called Lamb--Shift Hamiltonian $H_{\textnormal{LS}}$ represents a change of the system dynamics in comparison to the isolated system Hamiltonian $H_S$. $\Gamma$ has to be positive semidefinite. 
Note that both terms $\Gamma$ and $H_{\textnormal{LS}}$ vanish as the system--bath coupling goes to zero, thus retaining the von--Neumann equation of a closed quantum system.
}
Note that $\Gamma$ being positive semidefinite is a necessary and sufficient condition for the Lindblad equation to induce a completely positive dynamical map. Since complete positivity implies positivity but not vice--versa, a non--positive matrix $\Gamma$ is no indication on whether the master equation leads to a the violation of the positivity or not.

Following~\cite{hall2014canonical} we denote a master equation that has the same algebraic structure as the Lindblad form but has a non--positive semidefinite $\Gamma$ as having a \textbf{canonical form}. Every master equation that preserves trace and hermiticity has a representation in canonical form.

See Figure~\ref{fig:master_equations} in~\ref{sec:appendix_sec_derivation} for an overview of the discussed forms of master equations and their realizations using microscopic derivation.


\section{\label{sec:ERMEA}Energy resolved master equation approach}

After introducing the theoretical concept of a UDM which led to the Lindblad equation and before discussing several types of Born--Markov master equations we now introduce the basic idea of the energy resolved master equation approach, namely a physically motivated constructive way to represent the action of the superoperator in a small basis such that the determination of the steady state or relevant dynamics is possible without evaluating the full superoperator. This successive construction of the representation of the superoperator is monitored using convergence parameters for trace (Con.~1), hermiticity (Con.~2) and (complete) positivity (Con.~3) which help to decide on when to stop the construction.

This scheme provides a considerable and systematic improvement on the calculability of steady state dynamics in the context of many--body Lindblad and master equations for large Hilbert spaces.

To quote an example, we show in Sec.~\ref{sec:num_res_ERMEA} that reasonable effective representations of superoperators for six sites with this method can amount to a dimension of the order $10^3$. This should be compared to the Hilbert space of the full superoperator which has a dimension of $10^5$ to $10^6$ (compare table~\ref{tab:sigma_size}). 

\subsection{Reduction of the number $\mathcal{N}$ of considered matrix elements of $\sigma$ in a linear master equation} \label{sec:reduction_of_variables}
The Lindblad equation and most master equations can be written in terms of a linear generator $\mathcal{K}$ which represents a linear system of differential equations in the elements of the reduced density operator $\sigma$. 

Here, it is convenient to express the reduced density operator in the eigenbasis of the system Hamiltonian with eigenvectors $\ket{a}$ and eigenenergies $E_a$, since this basis allows to understand the equilibrium properties of the system and how they will change in the non--equilibrium situation. 

In this eigenbasis representation ($\sigma_{ab} = \bra{a} \sigma \ket{b}$) the master equation reads: $$\dot{\sigma}_{ab} = \sum_{cd}\mathcal{K}_{ab,cd} \sigma_{cd},$$
and can be cast in matrix form:
$$ \dot{\sigma}_i = \sum_{j \in\mathcal{I}} \mathcal{K}_{ij} \sigma_j,$$
 by choosing an adequate vectorization of the reduced density operator where we use the compound index $i=(a,b)$ and compound index set $\mathcal{I}$. This notation is used with a vectorization according to the columns (see also~\ref{app:choi}).

\begin{figure} 
	\begin{center}
		\includegraphics[width=0.8\textwidth]{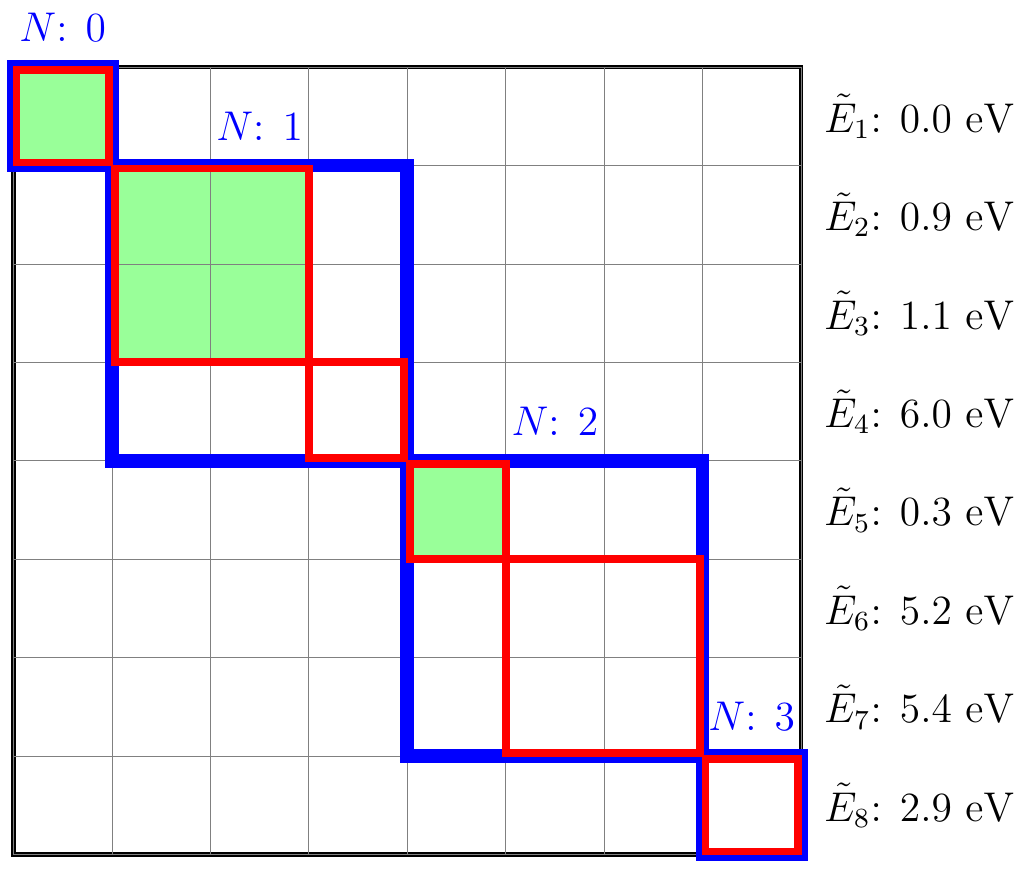}
		\caption{Illustration of the reduced density matrix $\sigma$ with dimension $n=8$ of a spinless three site system ($l=3$) and the three types of reduction of considered matrix elements $\mathcal{N}$. The matrix is given in the eigenbasis of the system with the grandcanonical energies $\tilde{E}$ indicated on the right side. They are sorted according to number of particles $N$. The blue squares mark the used matrix elements ($\mathcal{N}_N = 20$) in the first reduction exploiting the conservation of quantum numbers (number of particles). The red squares mark the second reduction of considered matrix elements (partial secular approximation) allowing coherences with an energy gap of $\Delta E = 1$ eV leading to $\mathcal{N}_{N,\Delta E} = 12$ matrix elements.
		The green area mark the used matrix elements in the third reduction ($\mathcal{N}_{N,\Delta E, }(\mathcal{E}) = 6$) using the ERMEA approach with a threshold of $\mathcal{E} = 1.1$~eV, leading also to a smaller dimension of the reduced density matrix $n(\mathcal{E}) = 4$.}
		\label{fig:reduction_of_variables}
	\end{center} 
\end{figure}

The superoperator $\mathcal{K}$ encodes the entire dynamics of the reduced density matrix, but in many cases not all of its information is needed or necessary for further calculations. 
We summarize three reasonable reductions of the number $\mathcal{N}$ of considered matrix elements which lead to a reduced number of differential equations and thus smaller generators:
\begin{enumerate}
	\item \textbf{Conserved quantum numbers:} If the full Hamiltonian $H$ conserves a quantum number (e.g. total particle number $N$ or $z$-component of the spin),  the differential equations for elements of the reduced density matrix which link states belonging to different quantum numbers (e.g. $\sigma_{a_N,b_{N'}}$) are decoupled from the differential equations for those elements belonging to the same quantum number (e.g. $\sigma_{a_N,b_N}$) and can be neglected in the master equation. The proof is given in~\ref{app:conserved} and the reduction is illustrated via the blue squares in figure~\ref{fig:reduction_of_variables}. 
	
	\item \textbf{Coherences in the system eigenbasis:} As pointed out in~\cite{mitchison2018non} coherences between different eigenstates do not arise in steady state in systems that are homogeneously coupled (equal coupling to all system sites) to baths in thermal equilibrium when using additive master equations, like the Born--Markov master equations that assume that there is no correlation between the baths. Coherences between different eigenstates of the systems can only arise from Born--Markov master equations if the system--bath coupling is non--homogeneous (e.g. each bath couples to a different system site) 
	which will be the case in the further discussion.
	
	Among them, those coherences between eigenstates which are separated by a rather large energy gap $\Delta E$, are related to a term of fast oscillations $e^{i\Delta E t}$ and as such lead to a decoupling of matrix elements in the master equation. The so--called secular approximation~\cite{barrat1961c,happer1972optical} or rotating wave approximation~\cite{whitney2008staying, dumcke1979proper} neglects all coherences between non--degenerate eigenstates in the derivation of the master equation. This is only valid if the energy gaps between eigenstates are large and thus oscillations are so fast that they can be averaged out. In recent years several approaches named \textbf{partial secular approximation}~\cite{jeske2015bloch, Farina2019coarse, cresser2017coarse, Cattaneo_2019} have been introduced for better control of this approximation.
	
	Since for large energy gaps the corresponding coherences can be neglected in the derivation due to decoupling arguments valid for the secular approximation, we introduce the partial secular approximation via a threshold energy gap $\Delta E$, which considers only coherences for eigenstates $\ket{a}$, $\ket{b}$ in the master equation with an energy difference below this threshold $\lvert E_a - E_b\rvert  \leq \Delta E$. See the red squares in figure~\ref{fig:reduction_of_variables}.
	\item \textbf{High energy states:} In many realistic setups where a weak--coupling description is meaningful, the occupation of the steady--state reduced density matrix $\sigma$ will be dominated by the energy landscape of the quantum system $S$ under the influence of the thermodynamic properties of the attached baths. This is exact in an equilibrium situation. 
	The dependence of the steady state on the energy landscape won't change significantly in a non--equilibrium situation when the system is coupled to several baths with different thermodynamic properties if they are not too different from the equilibrium parameters.
	
	 Especially eigenstates with a large grandcanonical energy won't contribute significantly to the steady state and can be neglected in many cases. This is the main idea of the ERMEA approach and will be discussed in the next subsection. For illustration of this reduction of considered matrix elements see also the green area in figure~\ref{fig:reduction_of_variables}.
\end{enumerate}

Note that the reduction of considered matrix elements $\mathcal{N}_X$ reduces also the dimension  of the linear generator $\mathcal{K}_X\colon \mathbb{C}^{\mathcal{N}_X} \rightarrow \mathbb{C}^{\mathcal{N}_X}$ where the domain / codomain is isomorphic to the Hilbert space of block--diagonal matrices $\mathbb{C}^{\mathcal{N}_X}\cong \mathcal{H}_X$. Here $X$ indicates the applied reductions (i) and (ii), compare with the examples in figure~\ref{fig:reduction_of_variables}.

To keep notation for the discussion of reduction (iii) simple and since ERMEA can be applied after  or without previous reductions we from now on omit the subscript for indicating the first two reductions.

Further note, that whereas the reductions (i) and (ii) only neglect off--diagonal matrix elements, reduction (iii) leads also to a shrinkage of the number of diagonal matrix elements $n(\mathcal{E})$. 

A more detailed discussion of errors induced by the partial secular approximation and ERMEA and a visualization of the reduction of variables is given in \ref{app:discussion_of_errors}.

\subsection{Steady state physics and high grandcanonical energy states in the weak--coupling regime}
In order to illustrate the reduction (iii) we discuss the steady state physics in the weak coupling regime and why high energy states can be neglected in the steady state.

Starting from a situation characterized by the weak coupling of the system to a single bath with the inverse temperature $\beta$ and the chemical potential $\mu$, the steady state 
 has a distribution of eigenstates $\ket{a}$ following $\frac{1}{Z}e^{-\beta (E_a - \mu N_a) }$ with $Z$ the grand canonical partition sum, the eigenenergy $E_a$ and the particle number $N_a$ of the eigenstate $\ket{a}$. Due to the exponential term one may already find many negligible states in the steady state if the system grandcanonical eigenenergies vary sufficiently.


A non--equilibrium situation can be realized by coupling to baths $\alpha$ with different chemical potentials $\mu_\alpha$ or temperatures $\beta_\alpha$ which cause a deviation from the equilibrium distribution of state occupation. We expect the distribution of the non--equilibrium steady state to be related to the underlying energy landscape (see also the discussion in~\cite{mitchison2018non} for homogeneous coupling). 

One can illustrate this idea by computing the purity $P = \Tr{\sigma^2}$ of the reduced density operator $\sigma = \sum_{a = 1}^{n} p_a \ket{a} \bra{a}$ which amounts to $P=1$ for a pure state and is close to one for the equilibrium steady state at low temperatures. 
A purity of the steady state close to one supports the idea of neglecting high energy states and can thus act as indication when the ERMEA approach is promising.
The minimal value of the purity $P =\frac{1}{n} \ll 1$ is reached for a fully mixed reduced density operator characterized by a flat distribution of eigenstates $p_a = \frac{1}{n}$. This indicates the extreme situation of flattening the occupations of the whole energy landscape of the system. For most realistic systems we expect the purity of the steady state not to change dramatically in the non--equilibrium situation (see also figure~\ref{fig:convergence_superoperator}).

In order to separate the system eigenstates in those, which are physically relevant for the master equation of interest and in those which can be neglected, we introduce a ''cost function`` $\chi$ for system eigenstates $\ket{a}$:
\begin{align}
\chi\left(a\right): =  E_a - \overline{\mu}N_a,
\end{align} namely the grand canonical energy
according to an effective averaged chemical potential~$\overline{\mu}$:
\begin{align} \overline{\mu} &= \frac{\sum_\alpha V_\alpha \mu_\alpha}{\sum_\alpha V_\alpha},\\
\end{align} with the total coupling weights $V_\alpha$ representing the influence of the bath onto the system. For attached baths with identical densities of states the coupling weights are given by the total sum of hoppings from bath to the system: $V_\alpha = \sum_{\kappa} \lvert V_{\alpha\kappa} \rvert$ [see (\ref{equ:coupling})].

If the coupling weights are identical and the applied chemical potentials in the two attached leads have an opposite sign $\mu_L = - \mu_R$, then the effective chemical potential $\overline{\mu}$ amounts to zero and the order of eigenstates in the protocol is determined simply by their eigenenergies according to the system Hamiltonian which will be the case in the numerical benchmark in Sec.~\ref{sec:num_res_ERMEA}.

This cost function $\chi$ sorts the eigenstates according to their initially assumed relevance for the steady state and as such will be the basis for the ERMEA protocol to determine the order of the eigenstates that are considered in the construction of the superoperator.

\subsection{Consecutive construction of the superoperator}
When searching for the steady state $\sigma$ by solving $\mathcal{K} \sigma = 0$, our approach consists of reducing the number of differential equations in the master equation described by the superoperator $\mathcal{K}_{ij}$ via deleting those rows $i=(a_1,a_2)$ and columns $j=(a_3,a_4)$ which belong to an irrelevant eigenstate $a_k$.
Since full construction and subsequent deletion is not efficient, a consecutive construction of the superoperator according to the just introduced cost function $\chi$ is preferred.
More specifically the approximated superoperator $\mathcal{K}(\mathcal{E})\colon\mathcal{H}(\mathcal{E}) \rightarrow \mathcal{H}(\mathcal{E})$ in the ERMEA approach is defined as:
\begin{align} 
\mathcal{K}(\mathcal{E}) &:= (\mathcal{K}_{ij}),\, \textnormal{ with  } \, i,j \in \mathcal{I}(\mathcal{E}),\label{equ:K_ERMEA} \\
\mathcal{I}(\mathcal{E}) &:= \{i \in \mathcal{I} \mid i=(a,b): \chi(a) < \mathcal{E} \land \chi(b) < \mathcal{E}\} \label{equ:index_set}, \\
 \mathcal{H}(\mathcal{E})&:= \textnormal{span}\left((A_i)\mid i \in \mathcal{I}(\mathcal{E}): i = (a,b), A_i = \ket{a}\bra{b} \right) ,
\end{align}
where $\mathcal{H}(\mathcal{E})$ is a subspace of the Hilbert space $\mathcal{H}$ of (block--diagonal) matrices and where $\mathcal{E}$ defines the \textbf{threshold} up to which eigenstates $\ket{a},\ket{b}$ are considered in the construction according to the cost function $\chi$. Increasing the threshold corresponds to adding more (deleting less) rows and columns to (of) the superoperator. 

\subsubsection{ERMEA and the Lindblad equation} \label{sec:second_threshold} ERMEA can be easily applied to Lindblad-like equations which use the following eigenbasis operators $A_i = \ket{a}\bra{b}$ in (\ref{equ:lindblad_equation})\footnote{Or more generally ladder operators with the property $[H_S,A_i] = c_i A_i$ for $c_i\in \mathbb{R}$.} as traceless orthonormal operators according to the Hilbert--Schmidt inner product.

For Lindblad equations using this eigenbasis operators and a column--wise vectorization of the reduced density operator (see also~\ref{app:choi}), the matrix of the full generator without applied reductions can be expressed as
\begin{align}
\mathcal{K} =& -\rmi\left(\mathds{1} \otimes (H_S + H_{\textnormal{LS}})^\transpose - (H_S + H_{\textnormal{LS}}) \otimes \mathds{1}\right) \label{equ:evaluation}\\
& +\left(C(\Gamma) - \frac{1}{2}[\mathds{1} \otimes (\textnormal{Tr}_{13} \Gamma)^\transpose +  (\textnormal{Tr}_{13}\Gamma) \otimes \mathds{1}] \right), \notag
\end{align}
where the Hamiltonians $H_S$ and $H_{\textnormal{LS}}$ are given in the matrix representation of the chosen basis states $(H_S)_{ab} = \bra{a} H_S \ket{b}$, $C(\Gamma)$ being the Choi matrix of Gamma matrix~$\Gamma_{abcd}$ (see~\ref{app:choi}), $(\textnormal{Tr}_{13} \Gamma)_{bd} = \sum_a \Gamma_{abad}$ the partial trace over two indices of $\Gamma$ and $^\transpose$~the transposition operator.

An interesting term which needs further discussion is the partial trace term $\textnormal{Tr}_{13} \Gamma$ since it requires the summation over all eigenstates and thus the knowledge of the full matrix $\Gamma$. 
Since a full construction of the matrix $\Gamma$ is computationally demanding, we introduce a second fixed threshold\footnote{This just introduced second threshold $\mathcal{E}_2$ is chosen to be fixed and much larger than the first threshold $\mathcal{E}$.}  $\mathcal{E}_2\gg \mathcal{E}$ up to which the partial trace of the matrix $\Gamma$ is performed in the ERMEA approach. We do this by assuming that terms which are ranked irrelevant according to the cost function $\chi$ will give only small contributions to the partial trace, since in most microscopic derivations the entries $\Gamma_{abcd}$ tend to zero for large energy differences $\lvert E_a - E_b\rvert$ and $\lvert E_c - E_d\rvert$.
 
This procedure is at least consistent since cutting rows and columns of a positive $\Gamma$ preserves its positivity.

In the next subsection we will discuss the fundamental question of how to distinguish between relevant and irrelevant states after having sorted them - or stated differently - the question of how to assess the convergence of ERMEA for the steady state with increasing $\mathcal{E}$.

\subsection{Convergence of ERMEA for the steady state}
After having defined the order of relevant states, the fundamental question is, how convergence of the steady state obtained with the approximated superoperator $K(\mathcal{E})$ [see (\ref{equ:K_ERMEA})] with increasing $\mathcal{E}$ can be proven and validated.

For the discussion of convergence we interpret the approximated generator $\mathcal{K}(\mathcal{E})$  as submatrix of the full generator $\mathcal{K}$ (respectively the generator after reductions (i) and (ii)) and use the following statement: 
\paragraph{Theorem 3.1.} We assume that the given master equation preserves trace, hermiticity and positivity and allows for a unique steady state, so that its generator $\mathcal{K}$ has a unique eigenvector $v^{(0)}$ to the eigenvalue zero. Then the submatrix $\mathcal{K}(\mathcal{E})$ has an eigenvalue zero if and only if the corresponding eigenvector reproduces the steady state $v^{(0)}$ of the full generator $\mathcal{K}$ via embedding.

\begin{proof}[Proof by contradiction]
	Given that the submatrix $\mathcal{K}(\mathcal{E})$ has a unique steady state $v^{(0)}$ (eigenvector $v^{(0)}$ to the eigenvalue zero), we assume that the embedding of $v^{(0)}$ in the full space by adding zeros yielding $\sigma^{(0)}$ does not reproduce the steady state of the full generator:
	$$\mathcal{K} \sigma^{(0)} = \begin{pmatrix} \mathcal{K}(\mathcal{E}) & B \\ C & D \end{pmatrix} \begin{pmatrix}
		\vec{v}^{(0)} \\ 0 
	\end{pmatrix} = \begin{pmatrix}
	0 \\ \vec{x}
\end{pmatrix},$$
with $\vec{x} = C\vec{v}^{(0)}\neq 0$. Rewriting to a density matrix structure according to the superindex set $\mathcal{I}(\mathcal{E})$ yields $\sigma^{(0)} = \begin{pmatrix} v^{(0)} & 0 \\ 0 & 0 \end{pmatrix}$ and $x =  \begin{pmatrix} 0 & x_c \\ x_c^\dagger & x_d \end{pmatrix}$.

Then the density matrix at an infinitesimal later time $\Delta t> 0$ is approximately $\sigma^{(0)}(\Delta t) = \begin{pmatrix} v^{(0)} & \Delta t x_c \\ \Delta t x_c^\dagger & \Delta t x_d \end{pmatrix}$. Since the full master equation is assumed to be trace preserving, the trace of $x_d$ has to be zero. Now due to the positivity preserving property of the full master equation the diagonal of $x_d$ has to be zero and due to Silvester's criterion it follows from the zero diagonal of $x_d$ that $x_c = x_d = 0$, thus contradicting the assumption $x \neq 0$. Thus $\sigma^{(0)}$ is the unique steady state of the full master equation.

	
\end{proof}

We conclude that an eigenvalue zero can only be found in the approximated generator if and only if it correctly reproduces the steady state. This motivates the definition of the following qualification factor for convergence of the steady state (\textbf{convergence number}) $\nu_c$ via the minimum of absolute value of the eigenvalues of $\mathcal{K}(\mathcal{E})$:
\begin{align}
\nu_c(\mathcal{E}) := \min_{\lambda \in \sigma\left(\mathcal{K}(\mathcal{E})\right)} \lvert \lambda \rvert. \label{equ:convergence_number}
\end{align}

\subsection{Validation of the approximated superoperator to generate a dynamical map}
Having shown the criterion of convergence of the steady state in the ERMEA approach we now introduce measures to validate and examine the approximated generator to see if it induces a trace, hermiticity and positivity preserving dynamical map $\mathcal{U}$. This is especially relevant for time evolution simulations close to steady state.

The analysis can be done within two frameworks, namely in the \textbf{picture of the time evolution} on the level of the dynamical map or in the \textbf{differential picture} on the level of the master equation that induces the dynamical map.

On the dynamical map level there is a general theory  that makes statements about the three mentioned properties of an arbitrary superoperator $\mathcal{T}\colon \mathbb{C}^n\otimes \mathbb{C}^n \rightarrow \mathbb{C}^m\otimes \mathbb{C}^m$, namely the well--known Choi theorem~\cite{choi1975completely,havel2003robust} based on the Choi matrix (or dynamical matrix~\cite{bengtsson2017geometry}) which is defined via 
\begin{align*}
C(\mathcal{T}):= \sum_{i,j \in \{1,\dots,n\}}  E_{ij} \otimes \mathcal{T}(E_{ij})\quad  \in \mathbb{C}^{nm\times nm} ,
\end{align*} where $E_{ij} = \ket{i}\bra{j}$ are orthonormal basis vectors of the space $\mathbb{C}^n\otimes\mathbb{C}^n$ with respect to the Hilbert--Schmidt inner product.

The theorem of Choi states that if and only if the Choi matrix $C(\mathcal{T})$ is positive, the superoperator $\mathcal{T}$ preserves positivity completely. Furthermore the trace is preserved if and only if the partial trace of the Choi matrix is equal to the identity matrix in the codomain: $\Tr_{13} C(\mathcal{T}) = \mathds{1}_m$.

Since the time integration of the generator (master equation) leading to the dynamical map and the evaluation of the corresponding Choi matrix may be expensive, we instead want to discuss directly on the generator level whether an approximated generator $\mathcal{K}(\mathcal{E})$ induces the demanded properties (positivity, trace and hermiticity preservation) in a dynamical map.

\subsubsection{Preservation of trace.} Applying ERMEA to a quantum master respectively Lindblad equation that induces a trace preserving dynamical map does in principle lead to a violation of trace preservation since 
in a rate equation picture, ERMEA does neglect transitions to some states classified as irrelevant, so those missing transitions weights can lead to a violation of trace preservation.

Since the dynamical map $\mathcal{U}$ induced by a generator $\mathcal{K}$ shall preserve the trace for all possible reduced density matrices, the time derivative of the trace or an arbitrary $\sigma$ has to be zero which leads to:
\begin{align}
0 &= \frac{\rmd}{\rmd t}\Tr{\sigma(t)} =  \Tr{ \mathcal{K} \sigma(t)} = \sum_{i \in \mathcal{I}_d} \sum_{j \in \mathcal{I}} \mathcal{K}_{ij} \sigma_j(t) \notag \\
&= \sum_{j\in \mathcal{I}} \sigma_j \left(\sum_{i\in \mathcal{I}_d} \mathcal{K}_{ij}\right)\Rightarrow \forall j: \sum_{i\in \mathcal{I}_d} \mathcal{K}_{ij} \overset{!}{=} 0,
\end{align}where $\mathcal{I}$ is the index set of all elements of the vectorized reduced density matrix and $\mathcal{I}_d$ the subset corresponding to the diagonal elements of the reduced density matrix.
This motivates the definition of the \textbf{trace number} $\nu_t$ for the dynamical map $\mathcal{U}(\mathcal{E})$ induced by the approximated generator $\mathcal{K}(\mathcal{E})$:
\begin{align}
\nu_t(\mathcal{E}) :=\frac{1}{\mathcal{N}(\mathcal{E}) \cdot n(\mathcal{E})} \sum_{j\in \mathcal{I}(\mathcal{E})} \left\vert \sum_{i\in \mathcal{I}_d(\mathcal{E})} \mathcal{K}_{ij}(\mathcal{E}) \right\vert,
\end{align} with the corresponding restricted index sets $\mathcal{I}(\mathcal{E})$ and $\mathcal{I}_d(\mathcal{E})$ with $\mathcal{N}(\mathcal{E})= \lvert \mathcal{I}(\mathcal{E})\rvert$ and $n(\mathcal{E}) = \lvert \mathcal{I}_d(\mathcal{E})\rvert$. A trace number $\nu_t(\mathcal{E})$ close to zero indicates convergence not only for the steady state but furthermore is a convergence measure for the approximated superoperator $\mathcal{K}(\mathcal{E})$ acting on the whole subspace $\mathcal{H}(\mathcal{E})$.

For many applications it is convenient to measure trace preservation not for the whole subspace in which the superoperator was approximated but only for the subspace which is relevant for the simulation. Thus the \textbf{partial trace number} $\tilde{\nu}_t$ measures trace preservation of the approximated generator $\mathcal{K}(\mathcal{E})$ acting on a subspace $\mathcal{H}(E) \subset \mathcal{H}(\mathcal{E})$ using a subthreshold $E < \mathcal{E}$:

\begin{align}
\tilde{\nu}_t(\mathcal{E}, E):=&  \frac{1}{\mathcal{N}(E) \cdot n(E)} \sum_{j\in \mathcal{I}(E)} \left\vert \sum_{i\in \mathcal{I}_d(E)} \mathcal{K}_{ij}(\mathcal{E}) \right\vert, \label{equ:partial_trace_number}
\end{align}
This number indicates how well an approximated superoperator preserves the trace for simulations of time evolution close to the steady state and provides interesting insights in the approximation property of ERMEA as will be discussed in Sec.~\ref{sec:num_res_ERMEA}.

 
\subsubsection{Preservation of hermiticity:} Since a dynamical map inherits the property of preserving hermiticity from its generator $\mathcal{K}$, we check the capability of the generator to fulfill Con.~2:
\begin{align*}
    \dot{\sigma}_{ij} &\overset{!}{=} \dot{\sigma}^*_{ji}, \\
    \sum_{lm} \mathcal{K}_{ij,lm} \sigma_{lm }  &\overset{!}{=}  \left(\sum_{ml} \mathcal{K}_{ji,ml} \sigma_{ml} \right)^* = \sum_{lm} \mathcal{K}^*_{ji,ml} \sigma_{lm}.
\end{align*}
Because this has to be valid for all possible Hermitian density operators $\sigma$ the following relation has to hold:
\begin{align}
    \mathcal{K}_{ij,lm} \overset{!}{=} \mathcal{K}_{ji, ml}^*. \label{equ:K_hermitian}
\end{align}
By comparison with the Choi matrix of the generator one can conclude (see~\ref{app:choi}) that the generator $\mathcal{K}$ preserves hermiticity if $C(\mathcal{K})$ is Hermitian.

We define the \textbf{hermiticity number} $\nu_h$ as the Frobenius norm of the anti-Hermitian part of $C(\mathcal{K(\mathcal{E})})$:
\begin{align}
 \nu_h(\mathcal{E}) = \sqrt{\sum_{ij\in \mathcal{I}(\mathcal{E})} |A_{ij}(\mathcal{E})|^2}, \quad A(\mathcal{E}) = \frac{C(\mathcal{K}(\mathcal{E}))-C(\mathcal{K}(\mathcal{E}))^\dag}{2}.
\end{align}
Since ERMEA does not violate the above defined relation~(\ref{equ:K_hermitian}) by deleting rows and columns, hermiticity will be preserved for any master equation which has this property.
\subsubsection{Preservation of positivity}\label{subsec:preservation_positivity}
The required (complete) positivity of the time evolution is equivalent to the positivity of the integrated master equation, e.g. the exponential term $\exp(\mathcal{K}\Delta t)$ in the time--independent case:
$ \sigma(t) = \rme^{\mathcal{K}(t-t_0)}\sigma(t_0)$.
In order to qualify the complete positivity of this exponential term one could use Choi's theorem as described above and check whether the whole spectrum of the Choi matrix $C\big(\exp(\mathcal{K}\Delta t)\big)$ is positive. 

Since any master equation that preserves trace and hermiticity can be brought into a canonical (Lindblad--like) form characterized by a matrix $\Gamma$ [see (\ref{equ:lindblad_equation})], one can measure the possible violation of complete positivity  directly from the negative eigenvalues of $\Gamma$. The sum of negative eigenvalues $\lambda_k^{(\mathcal{E})}$ of $\Gamma(\mathcal{E})$ can be used as qualification factor of complete positivity:
\begin{align}
\nu_m(\mathcal{E}):= \sum_{k=1}^{\mathcal{N}(\mathcal{E})} \left(\lvert \lambda_k^{(\mathcal{E})}\rvert - \lambda_k^{(\mathcal{E})}\right) = -2 \sum_{\lambda_k^{(\mathcal{E})} < 0} \lambda_k^{(\mathcal{E})}.
\end{align}
Hall et al.~\cite{hall2014canonical} used this sum of negative decoherence rates as measure of non--Markovianity and showed that for trace preserving dynamical maps it is equal to another measure of non--Markovianity, namely the trace norm rate of change of the Choi--transformed dynamical map~\cite{rivas2010entanglement}:
\begin{align} \nu_{mt}(\mathcal{E}) := \lim_{\Delta t \rightarrow 0} \frac{\lVert C(\mathds{1}(\mathcal{E}) + 	\Delta t \mathcal{K}(\mathcal{E}) )\rVert_1 - n(\mathcal{E}) }{\Delta t}, 
\end{align} with the identity matrix  $\mathds{1}_{\mathcal{E}}$ in the Hilbert space $\mathcal{H}(\mathcal{E}) \otimes \mathcal{H}(\mathcal{E})$ and  the dimension $n(\mathcal{E})$ of the block--diagonal matrices $\sigma \in \mathcal{H}(\mathcal{E})$.
The difference of these two measures yields a measure of trace preservation:
\begin{align}
\nu_{tc}(\mathcal{E}) := \lim_{\Delta t\rightarrow 0} \frac{n(\mathcal{E}) - \Tr\left\{C(\mathds{1}(\mathcal{E}) + \Delta t \mathcal{K}(\mathcal{E}))\right\} }{\Delta t} = \nu_m(\mathcal{E}) - \nu_{mt}(\mathcal{E}).
\end{align}
This relations are especially useful to evaluate the violation of complete positivity in the case of master equations which are not given in canonical form, e.g. the (not canonical) Redfield--Bloch master equations.

Since many master equations, like the Redfield--Bloch master equations, are known to violate complete positivity but in some setups preserve positivity, it may be of interest to examine positivity directly.
Also in the context of entanglement witnesses, the efficient determination of positivity for superoperators is  still a research topic~\cite{pastuszak2020quantifier}.
As shown in~\cite[Th.~4.4]{rivas2012open} or~\cite{kossakowski1972quantum} it is sufficient and necessary for positivity that the time evolution preserves the trace and is a contraction 
\begin{align}
\lVert \exp(\mathcal{K}t) \rVert_{\textnormal{op}} \leq 1, \, \forall t > 0,
\end{align} with the operator norm defined as:
$$ \lVert A \rVert_{\textnormal{op}} := \sup_{\lVert v \rVert_1 \neq 0} \frac{\lVert Av\rVert_1}{\lVert v \rVert_1}, \quad \lVert v \rVert_1 = \textnormal{Tr}\{ \sqrt{v^\dag v} \}. $$
We define the positivity number $\nu_p(\mathcal{E},t)$ as measure of the violation of positivity:
\begin{align}
\nu_p(\mathcal{E},t) = \lVert \exp (\mathcal{K}(\mathcal{E}) t) \rVert_{op} - 1.
\end{align}


 \section{\label{sec:master_equations}Born--Markov master equation approaches}
 This section is dedicated to the most common microscopically derived Born--Markov master equations with respect to their properties regarding positivity and ability to treat coherences in the weak coupling limit. An overview of the microscopically derived Born--Markov master equations is given in Figure~\ref{fig:master_equations} in~\ref{sec:appendix_sec_derivation}.
 

A straightforward microscopic derivation of the quantum master equation with the Born--Markov approximation which is exact in the weak--coupling limit, does not necessarily lead to a positive map when the coupling is not small. In fact those \textbf{Redfield--Bloch} type master equations (also called Born--Markov master equations) allow for coherences of quasi--degenerate states and preserve positivity in some situations but violate complete positivity as will be shown later. We will discuss the different variants of Redfield--Bloch master equations in Sec.~\ref{subsec:variants_redfield_bloch} and introduce a canonical version which will allow the straightforward representation of the master equation in canonical form which is advantageous compared to the other types of Redfield--Bloch master equations, since in these cases a transformation to the canonical form is usually complicated~\cite{whitney2008staying}. Furthermore the canonical form helps to analyse the violation of complete--positivity in the ERMEA approximation (measure of non--Markovianity $\nu_m$) systematically as will be shown in Sec.~\ref{subsec:violation_of_complete_positivity}.

While on the one hand it is widely discussed in which cases the Redfield--Bloch master equation can be applied despite the fundamental positivity issue~\cite{whitney2008staying, jeske2015bloch, eastham2016bath}, there are many approaches to further develop the Redfield Bloch equation into a Lindblad equation by applying additional approximations or averaging techniques.

The most prominent approach is applying the so-called secular approximation (also called rotating wave approximation~\cite{Breuer2002}) which states that coherences between different system eigenenergies are decoupled from the relevant dynamics of master equation (populations) and thus can be neglected in the derivation of the master equation. The assumption for which the secular approximation is valid requires the eigenenergies to be sufficiently gapped, so that the argument of fast oscillations that are averaged out holds (see also (ii) in Sec.~\ref{sec:reduction_of_variables}). Therefore, only the coherences between degenerate eigenstates are taken into account. Though this leads to a master equation in Lindblad form - we refer to this master equation as \textbf{Davies--Lindblad} (DL) master equation - which preserves complete positivity, it fails to correctly describe coherences between quasi--degenerate states which arise naturally in any realistic molecule.

A second approach, known as the \textbf{singular coupling limit} (using a second Markov approximation)~\cite{ Breuer2002, Schaller2014}, averages the energies of quasi--degenerate eigenstates in the dissipative term to obtain a Lindblad form~\cite{schultz2009quantum}.

Both approaches are valid under certain conditions which in many applications are not truly verified, thus different approaches avoiding those approximations have been suggested, namely master equations derived by the \textbf{coarse graining technique}~\cite{schaller2008preservation, majenz2013coarse, mozgunov2020completely} and the \textbf{PERLind master equation}~\cite{kirvsanskas2018phenomenological}, that can be derived from the here presented \textbf{canonical Redfield--Bloch} (CRB) master equation by performing a geometric mean of the hermitian part of the bath contribution term in Sec.~\ref{sec:perlind}\footnote{A good summary about four of those mentioned Born--Markov master equations can be found in~\cite{potts2019introduction}.}.


\subsection{Variants of first Markov approximations}\label{subsec:variants_redfield_bloch}
Before specifying the coupling Hamiltonian $H_I$ to express the influence of the bath in terms of Green functions (see \ref{sec:appendix_decomposition}), we discuss in detail the first Markov approximation that leads to different types of the Redfield--Bloch master equation.
The standard microscopic derivation (see~\ref{sec:appendix_sec_derivation}) leads to the following  integro--differential equation which is exact for $t_0=-\infty$:
\begin{align}
    \dot{\sigma} (t) = &-\rmi\Trb{\mathcal{L}_0 \rho(t)]} -\underbrace{\int_{0}^\infty \Trb{ \mathcal{L}_I \rme^{-\rmi\mathcal{L}_0 \tau} \mathcal{L}_I\,  \rho(t-\tau)} \rmd\tau}_{=:\mathcal{D}}, \label{equ:integro_differential_equation}
\end{align}
with superoperators as defined in~\ref{app:superoperators}.


The next step requires to perform a Markov approximation which in many derivations is done in the interaction picture (note, that in our derivation we have already switched back to the Schrödinger picture).
Depending on how the Markov approximation is performed in the dissipative integral term $\mathcal{D}$, different variants of the Redfield--Bloch master equations $\mathcal{K}_0, \mathcal{K}_1$ and $\mathcal{K}_2$ are derived after implementing the Born approximation (see~\ref{sec:app_sub_born_approx}).

The zeroth variant $\mathcal{K}_0$ is gained by simply substituting $\rho(t-\tau)$ by $\rho(t)$ claiming that the density matrix is rather constant over time and that the bath correlation function is shaped like a delta peak which leads to the integral term
$$ \mathcal{D}_0 = \int_0^\infty \Trb{\mathcal{L}_I \rme^{-i\mathcal{L}_0 \tau} \mathcal{L}_I \rho(t)} \rmd\tau.$$ 
The resulting master equation corresponds to master equation $K_1$ in~\cite{dumcke1979proper}\footnote{The assumption for the interpretation of $K_1$ in~\cite{dumcke1979proper} is, that although there is a bracket for the partial trace the Liouville operators all act also on $\rho$.} and has some disadvantages in the evaluation since the exponential superoperator function acts on both, the interaction Hamiltonian and the density matrix.

A more practical approach is realized by approximating the time dependence of $\rho$ on $\tau$ by a time evolution with respect to the decoupled Hamiltonian $H_0$: $\rho(t-\tau) \approx \rme^{\rmi\mathcal{L}_0\tau}\rho(t)$. The approximation is to replace the full Hamiltonian $H$ by $H_0$.
In this manner we correct for the time shift by a unitary time evolution induced by the decoupled Hamiltonian $H_0$.
There are two inequivalent ways of applying this approximation in~(\ref{equ:integro_differential_equation}):
\begin{enumerate}
    \item Propagate before the interaction with the environment by substituting $$\rho(t-\tau) \approx \textcolor{blue}{\rme^{\rmi\mathcal{L}_0\tau}} \rho(t).$$ 
    \item Propagate after the interaction with the environment by substituting $$\mathcal{L}_I \rme^{-\rmi\mathcal{L}_0 \tau} \mathcal{L}_I \rho(t-\tau) \approx \textcolor{blue}{\rme^{\rmi\mathcal{L}_0 \tau}} \mathcal{L}_I \rme^{-\rmi\mathcal{L}_0 \tau} \mathcal{L}_I \rho(t).$$
\end{enumerate}

The first approach leads to a master equation where the effective time evolution acts on the second coupling Liouville operator $\mathcal{L}_I$ in the integral term:
\begin{align}
   \mathcal{D}_1:= & \int_{0}^{\infty} \Trb{ 
    \Big[ H_I,  \rme^{-\rmi H_0 \tau} [H_I, \textcolor{blue}{\rme^{\rmi H_0 \tau}\rho(t)\rme^{-\rmi H_0 \tau}}] \rme^{\rmi H_0\tau}\Big] } \rmd\tau \notag\\
=&\Trb{\int_0^\infty \Big[ H_I, [\underbrace{ \rme^{-\rmi H_0 \tau} H_I  \rme^{\rmi H_0 \tau}}_{=H_I(\tau)}, \rho(t)\Big] \rmd \tau } \\
    =&\Trb{ \int_{0}^\infty 
    \mathcal{L}_I \, \mathcal{L}_I(\tau) \rmd\tau \rho(t)}.
\end{align}
Here we use the shorthand notation $\mathcal{L}_I(\tau) = \mathcal{L}_{\exp(-i\mathcal{L}_0 \tau) H_I}$.
One can also obtain the resulting master equation $\mathcal{K}_1$ by using the projection technique~\cite[(90)]{rivas2012open} or by applying the Markov approximation ($\boldsymbol{\rho}(t-\tau) \approx \boldsymbol{\rho}(t)$ in the interaction picture~\cite[(3.118)]{Breuer2002}.

The second approach leads to a similar integral term except with the difference that the time evolution acts on the first coupling Liouville operator $\mathcal{L}_I$:
\begin{align}
 \mathcal{D}_2:= & -\int_{0}^{\infty} \Trb{ \textcolor{blue}{\rme^{\rmi H_0\tau}}
    \Big[  H_I, \rme^{-\rmi H_0 \tau} [H_I , \rho(t)] \rme^{\rmi H_0 \tau}\Big]\textcolor{blue}{\rme^{-\rmi H_0 \tau}}}\rmd\tau\notag \\
=&-\Trb{\int_0^\infty \Big[\underbrace{\rme^{\rmi H_0 \tau} H_I \rme^{-\rmi H_0 \tau}}_{=H_I(-\tau)} , [H_I, \rho(t)] \Big] \rmd \tau} \\
    =&-\Trb{ \int_{0}^\infty 
\mathcal{L}_I (-\tau) \mathcal{L}_I \rmd\tau\,\rho(t)} .
\end{align}
The resulting master equation $\mathcal{K}_2$ is also derived and discussed in the paper of D\"umcke and Spohn where it is referred to as $K_2$~\cite{dumcke1979proper}.

The basic idea leading to the \textbf{canonical Redfield--Bloch master equation} $\mathcal{K}_{\textnormal{CRB}}$ is to average the two last master equations:
\begin{align}
    \dot{\sigma} (t) = -\rmi\Trb{[H_0, \rho(t)]} + \frac{1}{2}(\mathcal{D}_1 + \mathcal{D}_2).
\end{align}
Note that applying the secular approximation at the three discussed types of Redfield--Bloch master equations $\mathcal{K}_0$, $\mathcal{K}_1$ and $\mathcal{K}_2$ leads to the same Davies--Lindblad master equation [see (\ref{equ:davies_lindblad})]. Details are given in \ref{sec:appendix_secBM_lindblad}.
We will now show, how this master equation $\mathcal{K}_{\textnormal{CRB}}$ can be brought in canonical form.

\subsection{Canonical Redfield--Bloch master equation} \label{sec:canonical_Redfield_Bloch}




The fermionic interaction Hamiltonian $H_I$ is chosen in its most basic version, coupling a system degree of freedom to the attached baths, with single--particle terms
\begin{align}
H_I &= \sum_{\mu} c_\mu^\dag \sum_{\alpha} V_{\alpha  \mu}^{\textnormal{*}} d_{\alpha} + \textnormal{h.c.} = \sum_{s\mu} s c_\mu^s \sum_{\alpha } V_{\alpha  \mu}^s d_{\alpha}^{\overline{s}}, \label{equ:coupling}\\
& s \in \{-1,1\},\quad  \overline{s} = -s, \quad c^s = \begin{cases} c & s = -1 \\ c^\dag & s = +1 \end{cases}, \notag
\end{align}
and fulfills the condition that spin and total particle number $N$ (system and bath)  remain preserved. $d_{\alpha}$, $d_{\alpha}^\dag$ are the creation and annihilation operators of an electron on the first site of bath $\alpha$ whereas $V_{\alpha  \mu}$ are the coupling constants. 
Note that by performing the Jordan--Wigner transform one can express this interaction Hamiltonian as sum of tensor products with respect to system and bath~\cite[Chapter 2.1]{Schaller2014} but the aim will be to express the influence of the bath in term of single particle Green functions thus keeping creation and annihilation operators in its form.
\newpage
After applying the Born approximation, separating the bath contribution, decomposing the bath contributions in terms of equilibrium Green functions (see \ref{sec:appendix_decomposition}) we can bring the canonical Redfield--Bloch master equation in canonical form, in terms of the system eigenvectors $\ket{a}, \ket{b}, \ket{c}, \ket{d}$ to define traceless operators $A_i = \ket{a}\bra{b}$ and $A_j = \ket{c} \bra{d}$ using the collective index $i=(a,b), j=(c,d)$:

\tboxit{Canonical Redfield--Bloch master equation using the system eigenbasis}{
	
	\begin{align}
	\dot{\sigma}(t) = & -\rmi[H_S + H_{\textnormal{LS}}, \sigma(t)] \notag \\
	&+ \sum_{abcd} \Gamma_{ab|cd}^{\textnormal{CRB}}\Big(\ket{a}\bra{b} \sigma(t) \ket{d}\bra{c}  - \frac{1}{2} \{\ket{d}\bra{c} \ket{a}\bra{b}, \sigma(t)\}\Big), \label{equ:typeC}
	\end{align}
	with the Gamma matrix
	\begin{align}
	\Gamma_{ab|cd}^{\textnormal{CRB}} = \sum_{\alpha }& \bra{a} \sum_\kappa c_\kappa^{\overline{s}} V_{\alpha \kappa }^{\overline{s}} \ket{b} \frac{O^s_{\alpha}(E_{ba}) + O^s_{\alpha}(E_{dc})}{2}  \bra{d} \sum_\mu V_{\alpha  \mu}^{s} c_\mu^{s} \ket{c}, \label{equ:Gamma_vector}
	\end{align}
	using energy differences $E_{ba} = E_b - E_a$. The value of $s = s(a,b) = s(c,d)$ in the above formula is determined by the eigenvectors $\ket{a}$ and $\ket{b}$ and their corresponding particle numbers $N_a$ and $N_b$ with $s = N_b - N_a = \pm 1$.  The creation and annihilation operators yield zero for other particle number differences.  
	The term $O_{\alpha}^s(\omega)$ describes up to a factor of $2\pi$ the density of occupied states at energy $\omega$ in bath $\alpha$ (density of states times Fermi distribution function):
	\begin{align}
	 O_{\alpha}^s(\omega):= \rmi(G^R_{\alpha}(s\omega) - G^A_{ \alpha}(s\omega)) \cdot f^{\overline{s}}(s\omega|\beta_\alpha, \mu_\alpha), \label{equ:occupation}
	\end{align}
	using retarded and advanced Green functions $G^R_\alpha$ and $G^A_\alpha$ and the generalized Fermi function $f^+(\omega|\beta_\alpha, \mu_\alpha) = \left[\exp(\beta_\alpha(\omega - \mu_\alpha)) + 1\right]^{-1}$, $f^- = 1-f^+$, the bath temperature $\beta_\alpha$ and the chemical potential $\mu_\alpha$.

	The Lamb shift Hamiltonian $H_{\textnormal{LS}}$ is given by
	\begin{align}
	H_{\textnormal{LS}} := &-\rmi\frac{1}{4}\sum_{s\mu\kappa}
	c_\mu^s \sigma_{\mu \kappa}^{s}(-\mathcal{L}_S)  \{c_\kappa^{\overline{s}}\} -\rmi\frac{1}{4}\sum_{s\mu\kappa}
	\sigma_{\mu \kappa}^{s}(\mathcal{L}_S) \{ c_\mu^s \} c_\kappa^{\overline{s}}, \label{equ:Lamb_shift}
	\end{align} with the anti-Hermitian term of the bath contributions
	
	\begin{align}
	\sigma^s_{\mu\kappa} (\omega) = -\frac{1}{\pi} \sum_{\alpha } V_{\alpha \mu}^s \mathcal{P}\int_{-\infty}^\infty \frac{\left(G_{\alpha}^R(\omega') - G_{\alpha}^A(\omega')\right) f^{\overline{s}}(\omega'|\beta_\alpha,\mu_\alpha)}{\omega-s\omega'} \rmd\omega' V_{\alpha \kappa}^{\overline{s}}, \label{equ:sigma_full}
	\end{align}
	and the principal value integral $\mathcal{P}\int$, see~\ref{sec:appendix_decomposition}.

}

\subsubsection{Violation of complete positivity of the canonical Redfield--Bloch master equation:} \label{subsec:violation_of_complete_positivity} The form of the matrix element $\Gamma_{ab|cd}^\alpha$ (\ref{equ:Gamma_vector}) has the structure $\Gamma_{ij}^\alpha = v_{i}^\alpha\cdot (w_{i}^\alpha+w_{j}^\alpha)\cdot (v^\alpha_{j})^*$ with vectors $v^\alpha$ and $w^\alpha$. The desired positive semidefiniteness of $\Gamma^\alpha$ is determined by the eigenvalues of the matrix $W^\alpha$, $W^\alpha_{ij} = (w^\alpha_i+w^\alpha_j)$. The non--zero eigenvalues are given by 
\begin{equation}
\lambda_{1,2}^\alpha = \sum_{i=1}^n \textnormal{Re}(w_i^\alpha) \pm \sqrt{n\sum_{i=1}^n |w_i^\alpha|^2 - \left(\sum_{i=1}^n \textnormal{Im} (w_i^\alpha)\right)^2}.
\end{equation}
While in the secular approximation the vector $w$ is constant for the blocks of relevant combinations of $i$ and $j$\footnote{Since in the secular approximation only coherences between degenerate eigenstates are allowed only constant differences of eigenenergies will arise} leading to one positive non-zero eigenvalue in each block and thus a guaranteed completely positive map, the canonical Redfield--Bloch master equation is proven to violate complete positivity in general.

\subsection{PERLind master equation}\label{sec:perlind}
A recent approach to arrive at a Lindblad master equation was named PERLind (phenomenological position and energy resolved Lindblad) approach~\cite{kirvsanskas2018phenomenological} which can be derived from the canonical Redfield--Bloch master equation by performing the geometric mean instead of the arithmetic mean of the two hermitian parts of the bath contribution (compare especially~\cite[Suppl. Mat. (S.10)]{ptaszynski2019thermodynamics} and~\cite[(88)]{potts2019introduction}) to arrive at a positive semidefinite matrix $\Gamma$:
\begin{align}
\Gamma_{ab|cd}^{\textnormal{PERLind}} :=  \sum_{\mu\kappa\alpha} \bra{a} c_\kappa^{\overline{s}} \ket{b} \sqrt{\gamma_{\alpha,\mu\kappa}^s(E_{ba}) \cdot \gamma_{\alpha,\mu\kappa}^s(E_{dc})} \bra{d} c_\mu^s \ket{c} , 
\end{align}with the bath parts $\gamma_{\alpha,\mu\kappa}^s(\omega)$ of $\gamma_{\mu\kappa}^{s}(\omega)$ as defined in (\ref{equ:gamma}) and (\ref{equ:gamma_Green}).

Note that the geometric mean is performed within the sum over the baths which preserves the additivity of the dissipators in the Redfield--Bloch master equations.

By separating the dependencies of the system degrees $\mu$ and $\kappa$ one can see that within this approach the geometric mean only applies to the spectral bath density terms:
\begin{align}
\Gamma_{ab|cd}^{\textnormal{PERLind}} := \sum_{\alpha }& \bra{a} \sum_\kappa c_\kappa^{\overline{s}} V_{\alpha \kappa }^{\overline{s}} \ket{b} \sqrt{O^s_{\alpha}(E_{ba}) \cdot O^s_{\alpha}(E_{dc})}  \bra{d} \sum_\mu V_{\alpha  \mu}^{s} c_\mu^{s} \ket{c}. \label{equ:gamma_perlind}
\end{align}
Since these terms are always positive, complete positivity is guaranteed within the PERLind approach.

Note that the Lamb--shift Hamiltonian $H_{\textnormal{LS}}$ in the PERLind approach~\cite[Suppl. Mat.]{ptaszynski2019thermodynamics} is identical to the one introduced in the canoncial Redfield--Bloch master equation. Notice however, that $H_{\textnormal{LS}}$ does not commute with $H_S$, in contrast to the claim in the supplementary material of~\cite{ptaszynski2019thermodynamics}. This would only hold if the Lamb--shift Hamiltonian were diagonal in the system eigenbasis which is in general not the case.

Furthermore, note that although the PERLind approach is convenient since it can take into account all possible coherences between eigenstates (in contrast to the Davies--Linblad approach) and provides a complete positive map it is a \textbf{phenomenological approach} since a physical justification for the geometric mean is missing.

 \section{\label{sec:num_res_ERMEA}Numerical benchmark of destructive quantum interference in a 6-site model}
For testing ERMEA we consider a fermionic six site model with quasi--degenerate eigenstates which features a destructive quantum interference (DQI) effect and which is computationally quite expensive when building the full superoperator $\mathcal{K}$.
\begin{figure}[ht]
    \centering
    \includegraphics[width = 0.8\textwidth]{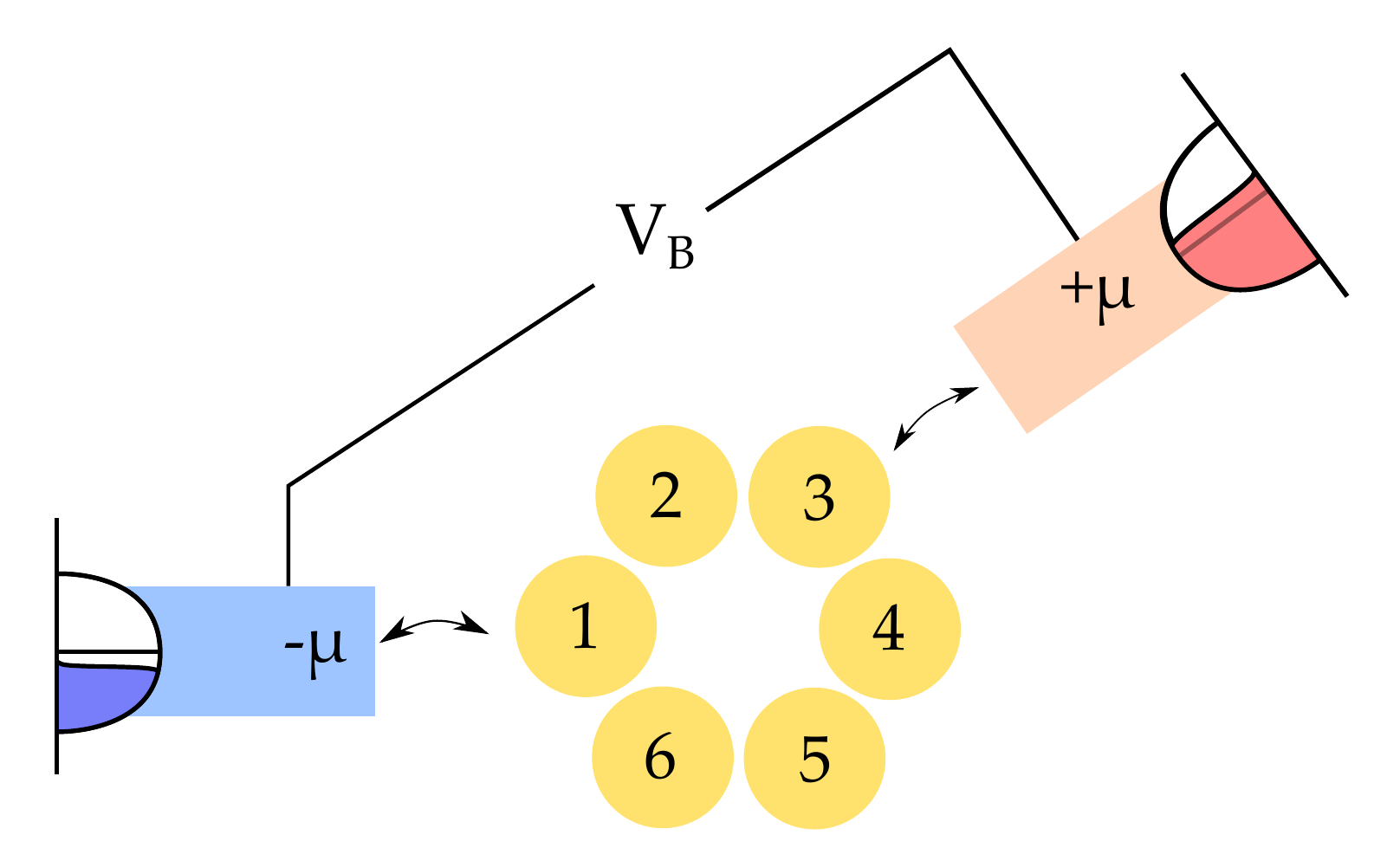}
    \caption{Sketch of the fermionic six site system coupled to two leads in the meta configuration.}
    \label{fig:system}
\end{figure}

\begin{figure}[ht]
    \centering
    \includegraphics[width = 0.8\textwidth]{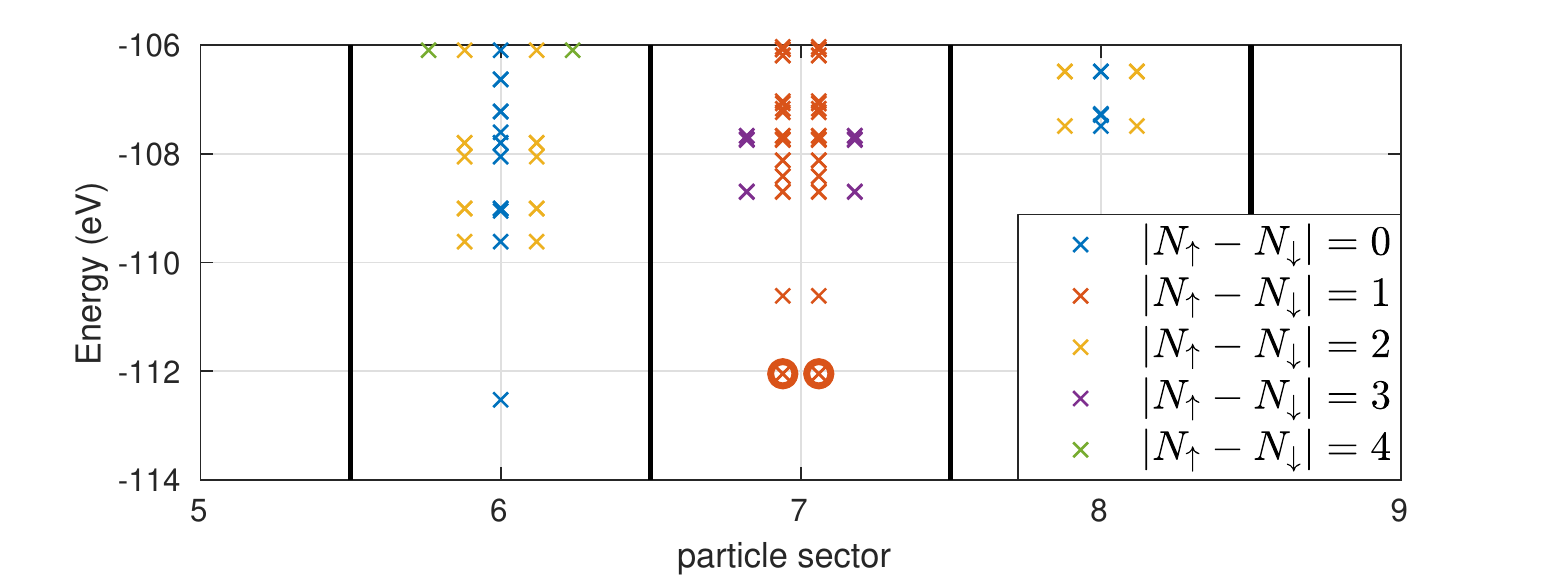}
    \caption{Low energy spectrum of the six site system separated by particle and spin number. The spectrum ranges from -113 to 0 eV. The quasi--degenerate eigenstates in particle sector seven are marked by circles.}
    \label{fig:energy_spectrum}
\end{figure}

The system Hamiltonian $H_S$ is motivated by DFT calculations of a D6h symmetric benzene molecule~\cite{rumetshofer2017first, DarauBegemannEtAl2009}, preserves particle number and spin and is given in energy units of electronvolts by
\begin{align}
 H_S =& \sum_{\sigma, ij} T_{ij}c_{i\sigma}^\dag c_{j\sigma} + \sum_{ij\sigma\sigma'} \frac{U_{i j}}{2}\left(n_{i\sigma} n_{j\sigma'} - \frac{n_{i\sigma} + n_{j\sigma'}}{2}\right) ,\notag\\
    T& = \begin{pmatrix} -3.8 & -2 & -0.3 & 0  & -0.3 & -2 \\ -2 & -3.8 & -2 & -0.3 &0 & -0.3 \\ -0.3 & -2 & -3.8 & -2 & -0.3 & 0\\0 & -0.3 & -2 & -3.8 & -2 & -0.3 \\ -0.3 & 0 & -0.3 & -2 & -3.8 & -2 \\ -2 & -0.3 & 0 & -0.3 & -2 & -3.8  
    \end{pmatrix}, \label{equ:Hamiltonian}\\
    U& = \begin{pmatrix} 8 & 5 & 3 & 2 & 3 & 5\\5 & 8 & 5 & 3 & 2 & 3 \\ 3 & 5 & 8 & 5 & 3 & 2 \\ 2 & 3 & 5 & 8 & 5 & 3 \\    3 & 2 & 3 & 5 & 8 & 5 \\    5 & 3 & 2 & 3 & 5 & \textbf{8.1} \notag
    \end{pmatrix}.
\end{align}
The small deviation in the Coulomb repulsion on site six breaks the symmetry and all  degeneracies of eigenstates in the same particle and spin sector vansish. Still some quasi--degenerate eigenstates with an energy gap $\Delta E < 0.1$ eV remain.
The spectrum of this fermionic six site-system is depicted in Fig~\ref{fig:energy_spectrum}. The first excited states (circle) in the seven particle sector are such quasi--degenerate eigenstates.

Attached to this six site model are two leads modelled by semi--infinite tight binding chains with an internal hopping of $t_B = 6 $ eV (bandwidth = 24 eV) and zero on-site energies at temperature $\beta = 20$. The coupling to the central system [Eq.~(\ref{equ:coupling})] is given by $V_{L,1} = V_{R,3} = 0.1$~eV. The left respectively right lead couples to site one respectively site three. This contact scheme is called meta configuration (see also figure~\ref{fig:system}) which gives rise to DQI~\cite{markussen2010relation}.

The non--equilibrium situation is obtained by applying a bias voltage $V_B$ on the leads by shifting the chemical potential $\mu_L = -\frac{V_B}{2}, \mu_R = \frac{V_B}{2}$ in the Fermi function. 

\subsection{Reduction of considered matrix elements in the master equation} The system under consideration preserves the total particle number and spin. Thus the system Hamiltonian $H_S$ is block-diagonal in these quantities which also holds for the bath Hamiltonian $H_B$. Since the coupling~[see~(\ref{equ:coupling})] preserves the total particle number and spin of system and bath, the reduced density matrix and its defining master equation will show the same block-diagonal structure [reduction (i)]. 

Applying the secular approximation restricts the reduced density matrix to be diagonal in its eigenenergies, since no degenerate eigenstates within the same particle and spin sector occur due to the small deviation of the Coulomb repulsion on site six. 

The following table~\ref{tab:sigma_size} compares the number of relevant elements of the density operator obtained from different reductions discussed in Sec.~\ref{sec:reduction_of_variables}. 
%
%
%
\begin{table}[ht]\caption{Effect of reduction of considered matrix elements on the dimension $\mathcal{N}$ of the superoperator $\mathcal{K}$ for the 6--site model. Within the ERMEA approach the dimension of the superoperator depends on (beside the energy tolerance $\Delta E $ of the partial--secular approximation) the applied bias voltage $V_B$ and the chosen tolerance $\delta$ for the convergence number $\nu_c$ (\ref{equ:convergence_number}). Compare also the illustration of the discussed reduction of matrix elements in figure~\ref{fig:reduction_of_variables}.}
\footnotesize

  \begin{tabular}{@{}lrr} \br
         \textbf{Reduction of considered matrix elements due to:} &  $\mathcal{N}$ & Color code in figure~\ref{fig:reduction_of_variables} \\ \mr
         No reduction, full density operator &                     $16\,777\,216$ & black\\ \mr
         Conserved particle and spin number &                     $853\,776$    & blue\\ \mr 
         Partial--secular approximation $\Delta E = 0.2$ eV &   $43\,572$ & red\\
         Partial--secular approximation $\Delta E = 0.1$ eV &   $17\,050$ &\\
         Secular approximation  $\Delta E = 0$ eV  &                 $4096$ & \\ \mr
         ERMEA, $V_B =  5$ V, part.--sec. approx. $\Delta E = 0.1$, $\delta = 10^{-6}$ & $5283$ & green\\ \br
    \end{tabular}
    
    \label{tab:sigma_size}
\end{table}

 If one neglects all coherences between different eigenenergies (secular approximation) -- especially the interactions between quasi--degenerate eigenenergies -- then as pointed out in~\cite{DarauBegemannEtAl2009} the destructive quantum interference is not obtained anymore (see figure~\ref{fig:current}). On the other hand, taking into account all coherences between different energy states (after exploiting the separation of dynamics due to conserved quantities) would produce a rather large superoperator matrix $\mathcal{K}$ which would make the search for the steady state computationally very expensive.  
 
 To take those interactions which are relevant into account and discard those which are not, it is convenient to pursue an intermediate approach of a \textbf{partial--secular approximation} which considers coherences between eigenenergies within a specified energy range~$\Delta E$. As discussed, the size of the generator can further be shrunken by using ERMEA (Sec.~\ref{sec:reduction_of_variables}) and strongly depends on the applied bias voltage $V_B$ as visualized in figure~\ref{fig:convergence_superoperator}.
 
 We now present results obtained by applying the partial secular approximation in the CRB and PERLind master equation with including coherences with an energy range of $\Delta E = 0.1$ eV. A comparative calculation with $\Delta E = 0.2$ eV didn't show a significant change in the current characteristics.
 The second threshold for virtual excitations $\mathcal{E}_2$ (see Sec.~\ref{sec:second_threshold}) was set sufficiently high to 80 eV above the groundstate energy.

\subsection{Convergence using ERMEA} We applied ERMEA on the following three master equations and calculated the steady state density matrix and the related current: 
\begin{itemize}
	\item the Davies--Lindblad (DL) master equation (\ref{equ:gamma_davies_lindblad}),
	\item the PERLind master equation (\ref{equ:gamma_perlind}) and
	\item the canonical Redfield--Bloch (CRB) master equation (\ref{equ:Gamma_vector}). 
\end{itemize}

\begin{figure}[h!]
	\centering
	\includegraphics[width = 1.0\textwidth]{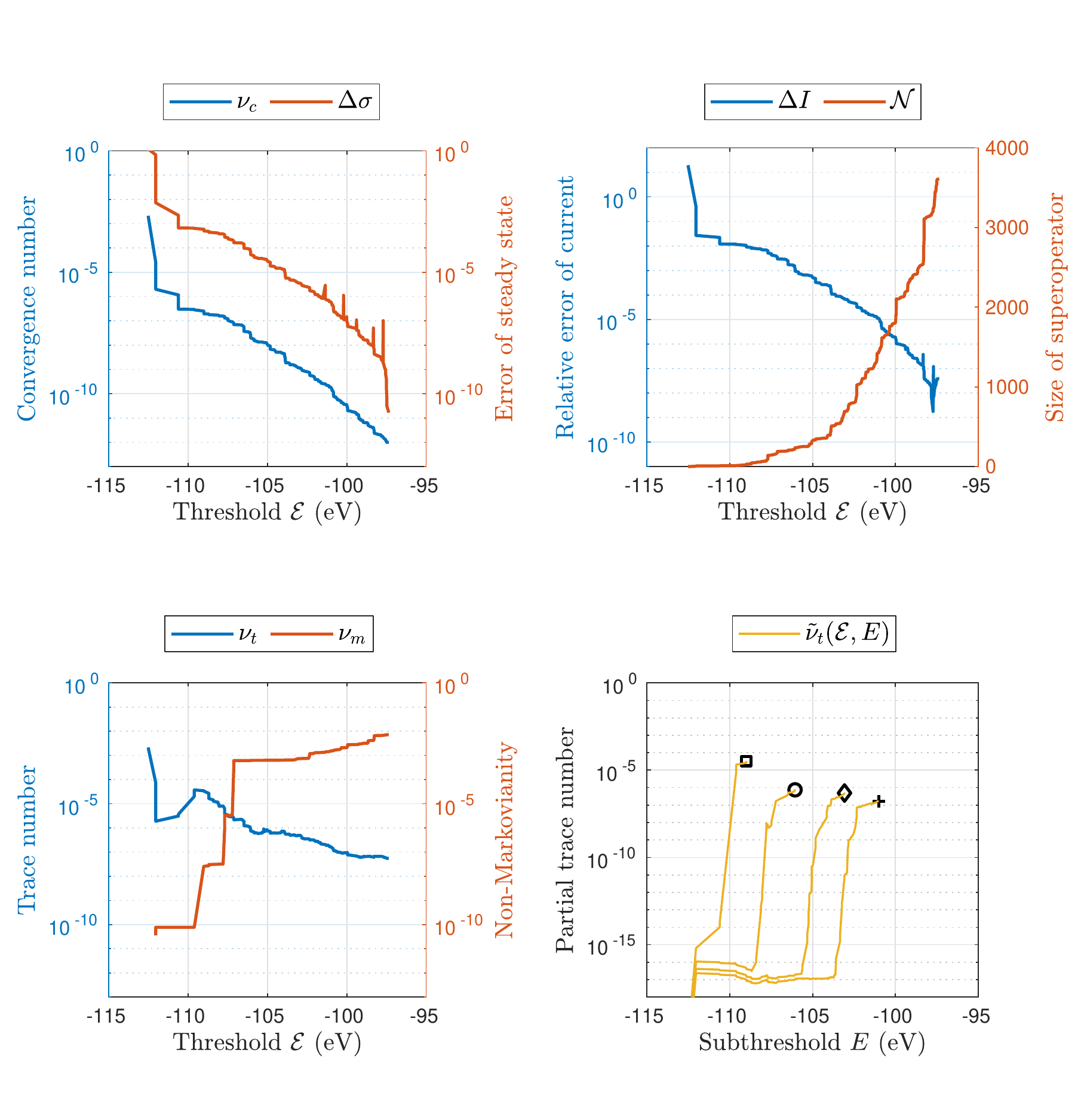}
	\caption{Convergence trend of approximated superoperator $\mathcal{K}_{\textnormal{CRB}}(\mathcal{E})$ for a fixed applied voltage of $V_B = 3.6$ V as a function of the grandcanonical energy threshold $\mathcal{E}$. 
		
		The upper left panel shows the convergence number $\nu_c$ and the convergence of the steady state $\Delta \sigma$ using the Frobenius norm of the difference from the converged steady state matrix which is exact due to theorem~(3.1).
		
		The upper right panel depicts the relative error of the calculated current $\Delta I$ and the dimension of the superoperator $\mathcal{N}$.
		
		The lower left panel shows the trace number $\nu_t$ and the Non-Markovianity $\nu_m$ which is a measure of the violation of complete positivity.
		
		In the lower right panel the partial trace numbers $\tilde{\nu}_t(\mathcal{E}_i,E)$ are presented for fixed energy thresholds $\mathcal{E}_i = -109$~eV (square), $-106$~eV (circle), $-103$~eV (diamond) and $-101$~eV (plus) as a function of the domain restriction $E$.}
	\label{fig:convergence_trend_fixed_Vb}
\end{figure}

\paragraph{Steady state convergence:}
Figure~\ref{fig:convergence_trend_fixed_Vb} shows the convergence and calculated quality factors of the superoperator $\mathcal{K}_{\textnormal{CRB}}$  for increasing grandcanonical energy threshold $\mathcal{E}$ at a fixed bias voltage of $V_B = 3.6$~V. It comes out clearly that the steady state convergence number $\nu_c$ as well as the steady state density matrix and the current through the system converge well with an increased threshold. Since the hermiticity remains preserved within ERMEA the hermiticity number $\nu_h$ is zero within numerical accuracy. 
\paragraph{Convergence for time evolution simulations:}
The trace preservation of the approximated superoperator shows an interesting behaviour which is relevant for applying ERMEA for time evolution simulations. Whereas the trace number $\nu_t$ - measuring the violation of trace preservation for the whole reduced space $\mathcal{H}(\mathcal{E})$ of the approximated superoperator $\mathcal{K}(\mathcal{E})$ - does not converge that fast, the partial trace number $\tilde{\nu}_{t}(\mathcal{E},E)$ (\ref{equ:partial_trace_number}) indicates that the trace for a subspace $\mathcal{H}(E) \subset \mathcal{H}(\mathcal{E})$ is preserved well. 

Four of these partial trace numbers are depicted in figure~\ref{fig:convergence_trend_fixed_Vb} and show that the trace preservation becomes worse for the higher energy terms a superoperator is acting on. The action of the approximated superoperator on high grandcanonical energy parts of a density matrix does not fulfill the trace preserving property. This is a clear artifact of the approximation since shifting the threshold $\mathcal{E}$ to higher energies repairs this failure for fixed eigenenergies $E$.

For simulations using time evolution both thresholds $\mathcal{E}$ and $E$ have to be chosen in such a way, that the expected states that might occur in the simulation lie within the subspace $\mathcal{H}(E)$ of preserved trace ($\tilde{\nu}_t(\mathcal{E},E) < \delta$).

\paragraph{Violation of complete positivity:} The lower left panel of figure~\ref{fig:convergence_trend_fixed_Vb} shows the violation of complete positivity ($\nu_m$) which is non--zero for the CRB master equation and increases with higher threshold $\mathcal{E}$ since more negative decoherence rates are considered in the construction of the approximated superoperator.

The PERLind and Davies--Lindblad approach yield the same results and trends except for zero violation of complete positivity and are therefore not plotted.

\begin{figure}[h!]
	\centering
	\includegraphics[width = 0.8\textwidth]{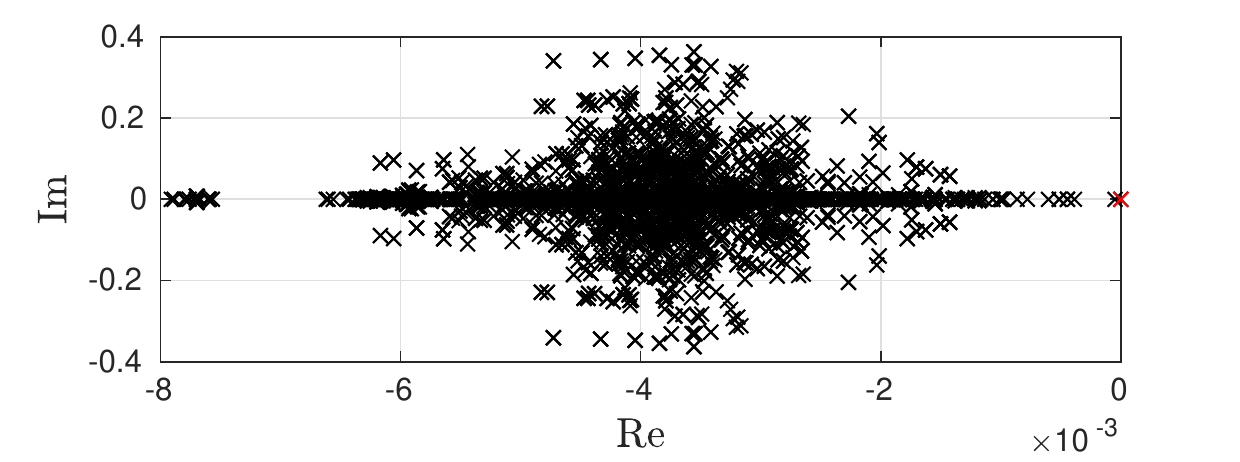}
	\caption{Spectrum of the eigenvalues of the converged superoperator $\mathcal{K}_{\textnormal{CRB}}(\mathcal{E}_{\delta_2})$ for an applied bias voltage of $V_B = 3.6$ V. The steady state eigenvector to eigenvalue zero is marked in red.}
	\label{fig:K_spectrum}
\end{figure}

Figure~\ref{fig:K_spectrum} illustrates the spectral properties of the converged superoperator $\mathcal{K}_{\textnormal{CRB}}$ for a bias voltage of $V_B = 3.6$ V. One can clearly see that all but one of the eigenvalues have a negative real part, which reflects the dissipative properties of the derived master equation.

\begin{figure}[h!]
    \centering
    \includegraphics[width = 0.8\textwidth]{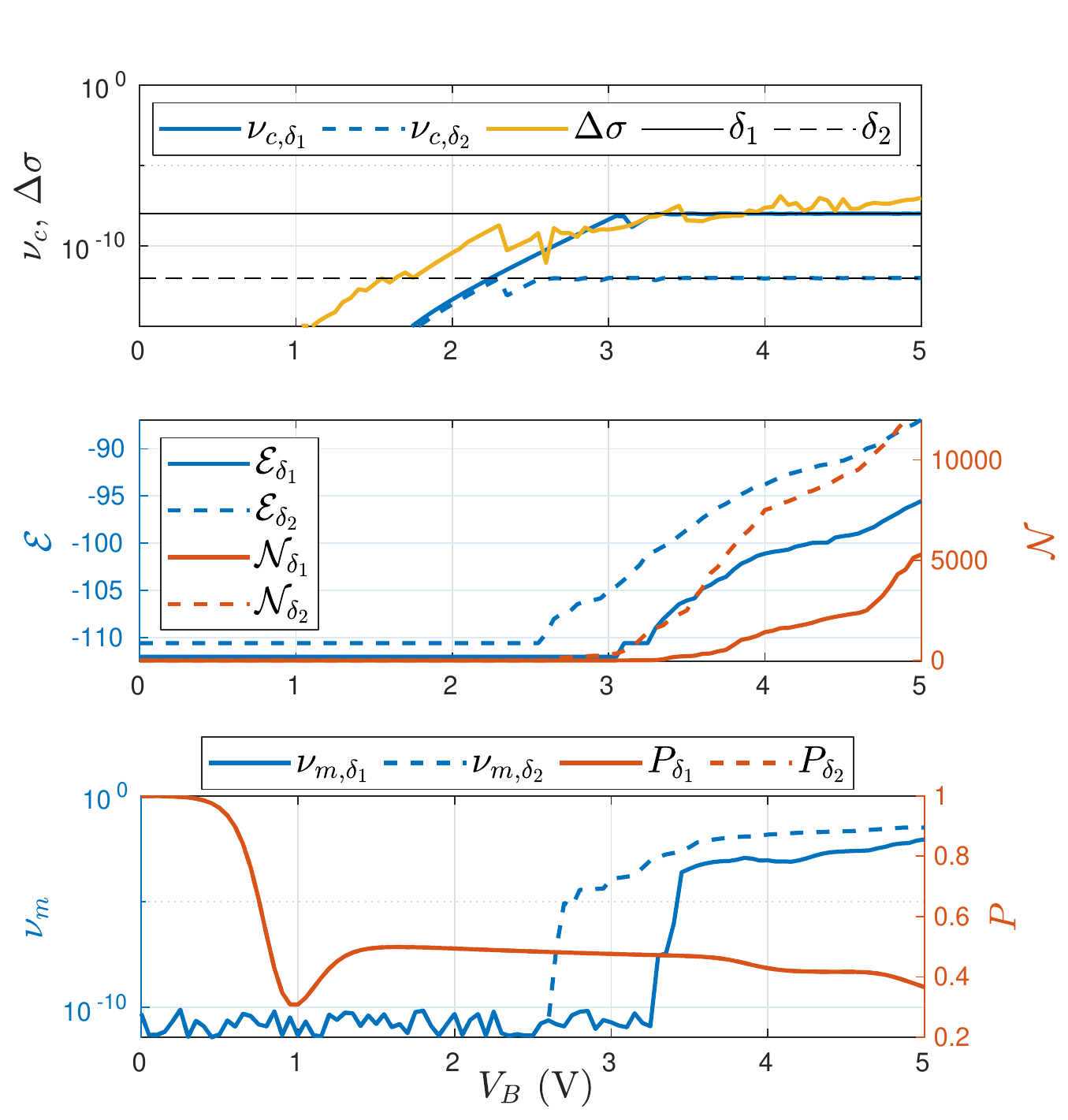}
    \caption{Monitoring of the application of ERMEA on $\mathcal{K}_{\textnormal{CRB}}$ using different convergence number tolerances $\delta_1 = 10^{-6}$ (solid lines) and $\delta_2 = 10^{-12}$ (dashed lines) for different applied bias voltages $V_B$.
    	
    Upper panel: Steady state convergence number $\nu_{c,\delta_i}$ obtained for different tolerances $\delta_1$ and $\delta_2$ (indicated with black solid and black dashed lines) as well as the deviation $\Delta\sigma$ of the obtained density matrix $\sigma_{\delta_1}$ from $\sigma_{\delta_2}$ using the Frobenius norm as function of applied bias voltage $V_B$.
    	
    Middle panel: Converged grandcanonical energy thresholds $\mathcal{E}_{\delta_1}$ and $\mathcal{E}_{\delta_2}$ (left axis, blue) and dimension of converged superoperator $\mathcal{N}$ (right axis, red) as a function of applied bias voltage $V_B$ for the two tolerances $\delta_1$ and $\delta_2$.
    	
    Lower panel: Violation of complete positivity $\nu_m$ (left axis) and purity $P$ of the steady state  (right axis) for the two tolerances as a function of applied bias voltage $V_B$. The curves of the obtained purities lie on top of each other.}
    \label{fig:convergence_superoperator}
\end{figure}

\paragraph{Convergence of ERMEA as function of applied bias voltage $V_B$:}
Figure~\ref{fig:convergence_superoperator} shows the computational effort required to reach convergence for different tolerances $\delta_1 = 10^{-6}$ and $\delta_2 = 10^{-12}$ of the convergence number $\nu_c$ as a function of applied bias voltage $V_B$.  Further $\Delta \sigma$ depicts the Frobenius norm of the difference of the steady state density matrices obtained with the tolerances $\delta_1$ and $\delta_2$. The final threshold $\mathcal{E}_{\delta_i}$ and the number of considered matrix elements $\mathcal{N}_{\delta_i}$ needed to obtain a convergence number below the indicated tolerance are shown in the middle panel. The last panel shows the purity of the steady state matrix and the violation of complete positivity (measure of non--Markovianity $\nu_m$) of the approximated superoperator $\mathcal{K}_{\textnormal{CRB}}$ over the bias voltage $V_B$.

\begin{figure}[h!]
    \centering
    \includegraphics[width = 0.8\textwidth]{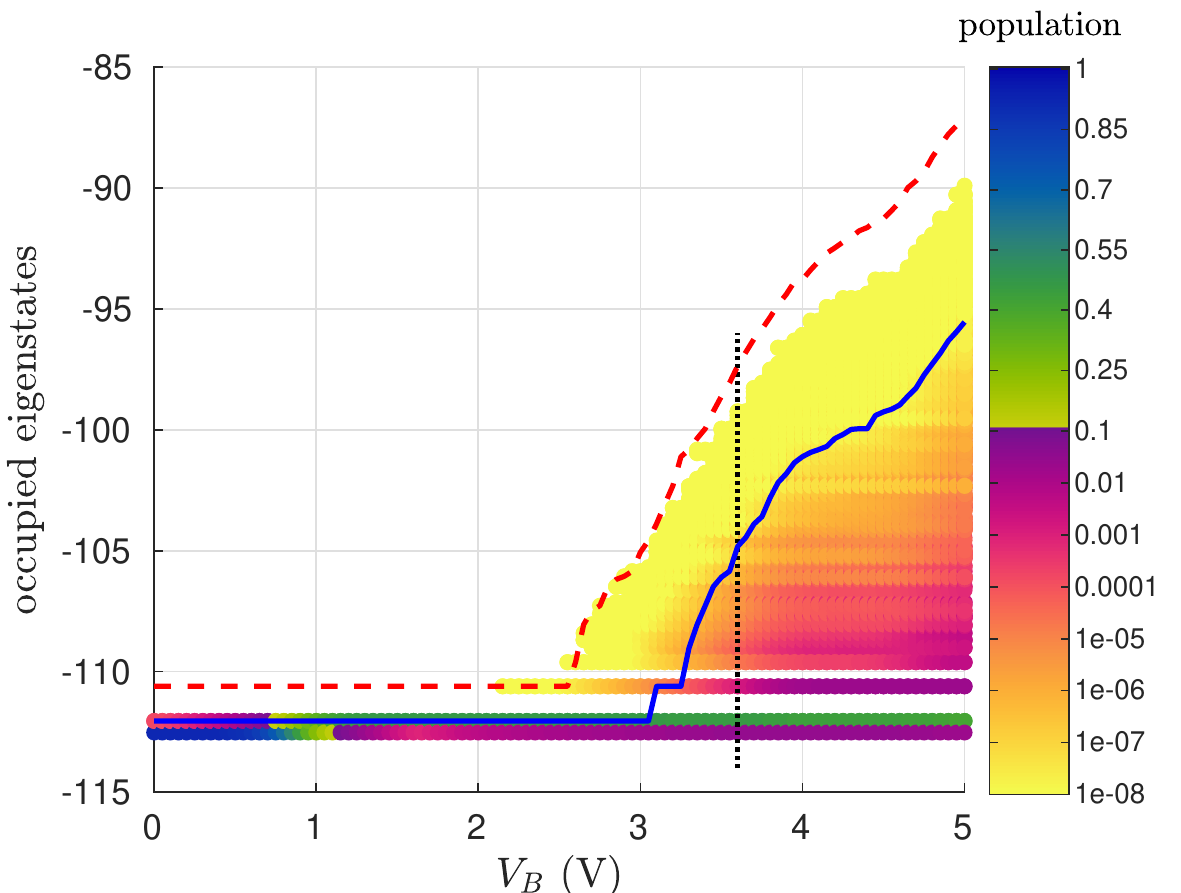}
    \caption{Distribution of entries of the steady state density matrix according to their eigenenergies as a function of applied bias voltage $V_B$. The steady state density matrix was obtained from $\mathcal{K}_{\textrm{CRB}}(\mathcal{E}_{\delta_2})$. The dashed and solid lines correspond to the grandcanonical energy thresholds $\mathcal{E}_\delta$ for which the convergence number $\nu_c$ was below the tolerances $\delta_1 = 10^{-6}$ (solid line) and $\delta_2 = 10^{-12}$ (dashed line). The dotted line marks the voltage $V_B$ for which the convergence trend of the approximated superoperator is discussed in figure~\ref{fig:convergence_trend_fixed_Vb}. }
    \label{fig:distribution}
\end{figure}

Figure~\ref{fig:distribution} shows the distribution of the steady state density matrix entries and the used grandcanonical energy thresholds $\mathcal{E}$ (dashed lines) to reach a convergence number $\nu_c$ below the tolerances $\delta_1 = 10^{-6}$ and $\delta_2 = 10^{-12}$ as a function of applied bias voltage $V_B$.   The first excited quasi--degenerate eigenstates (dark green) are the dominant states in the steady state situation over a wide voltage range.

\begin{figure}[ht!]
    \centering
    \includegraphics[width = 0.8\textwidth]{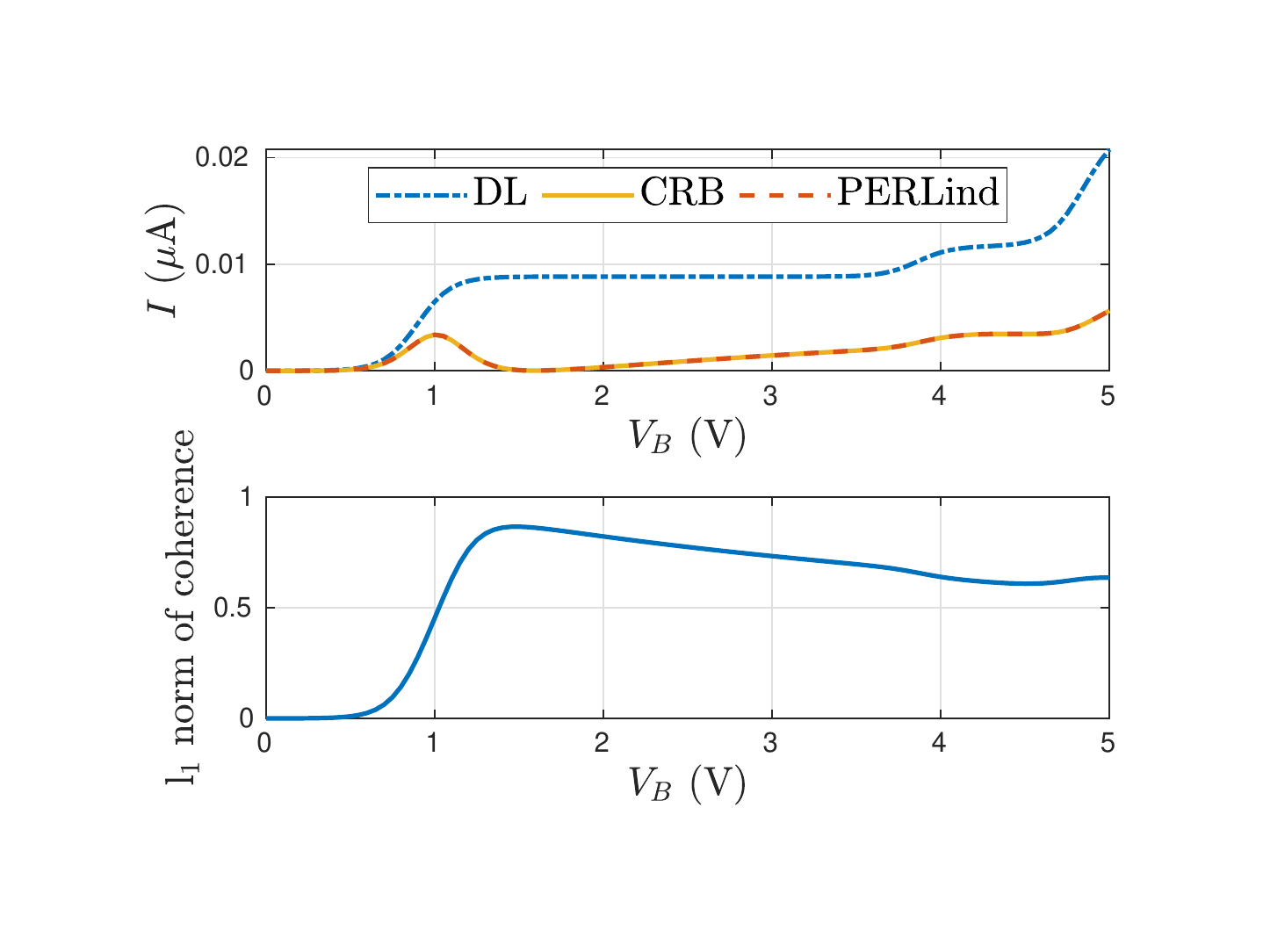}
    \caption{Upper panel: Current calculated for the six-site system the DL, CRB and PERLind master equations as a function of applied bias voltages $V_B$ using ERMEA with a convergence number tolerance $\delta_2 = 12^{-12}$. The occurring interference effects are not reproduced with the Davies--Lindblad master equation, because it neglects the quasi--degenerate states and thus cannot reproduce the rising coherences between those states~\cite{DarauBegemannEtAl2009}. 
    	
    	Lower panel: Amount of coherences defined as the $l_1$--norm of off--diagonal matrix elements of the resulting density matrix as a function of applied bias voltage $V_B$.}
    \label{fig:current}
\end{figure}

Finally we present the current in figure~\ref{fig:current} using the current formula derived in~\ref{sec:app_current} to emphasize the importance of taking quasi--degenerate eigenstates into account. These states build up strong correlations visible in the large off-diagonal entries of the steady state density matrix. The current was calculated as a function of the applied bias voltage $V_B$ for the three discussed master equations DL, CRB and PERLind. The secular approximation clearly fails to display the destructive quantum interference in this setup.




\section{\label{sec:conclusion}Conclusion}
This paper presents the energy resolved master equation as an approach to improve the computability of steady state density matrices of many--body open quantum systems with large Hilbert spaces. It may contribute to adopt more advanced master equation techniques dealing with larger systems. Especially the quality factors provide a useful tool on the decision of convergence of approximately calculated superoperators. We introduced a new variant of the Redfield--Bloch master equation which can be brought in canonical form and enables a better analysis of the violation of complete positivity. 
 
 It is striking to observe that although the Redfield--Bloch approach should be valid in the weak--coupling regime it fails to provide a valid universal dynamical map for quasi--degenerate eigenstates since complete positivity is always violated. This is because weak--coupling means $\lVert \Gamma \rVert \ll \Delta E$ which is not fulfilled by quasi--degenerate states. Since violation of complete positivity is a measure of non--Markovianity, this is a strong hint, that quasi--degenerate eigenstates in the system induce correlations between system and bath. 
 
 We thus regard the Redfield--Bloch master equations as strong tool to detect and analyse coherences between system and bath in open quantum systems. The developed program used in this paper will be released in near future under the name QUENTIN - QUantum ENtanglement and Transport INvestigator\footnote{https://github.com/dorn-gerhard/QUENTIN}.

 \ack 
 We thank M. Rumetshofer and M. Cattaneo for fruitful discussions.
 One of us (EA) gratefully acknowledges financial support by the Austrian Science Fund (FWF), grant No. P26508. The computational results presented have been achieved in part using the Vienna Scientific Cluster (VSC), project ID 71033.

\appendix
\section{Reduction of matrix elements} \label{app:conserved}
\subsection{Conserved quantities}
Assume that all three components of the Hamiltonian of the open quantum system have a conserved quantity such as the particle number (or spin). Then it holds that the total Hamiltonian, as well as the system and bath Hamiltonian commute with the particle number operator: $$ [H,\hat{N}] = [H_S,\hat{N}] =  [H_B, \hat{N}]= 0.$$
Thus the corresponding eigenvectors of the total system $H \ket{A^{(N)}} = E_A \ket{A^{(N)}}$ as well as the eigenvectors fo the subsystems $H_S \ket{a^{(N_S)}} = E_a \ket{a^{(N_S)}}$,  $H_B \ket{\alpha^{(N_B)}} = E_\alpha \ket{\alpha^{(N_B)}}$ belong to a distinct particle numbers $N$ respectively $N_S$ and $N_B$.
The von--Neumann equation for the full density operator $\rho$ shows that due to the conserved particle number the dynamics of the full system are separated with respect to the particle numbers:
\begin{align*}
i\dot{\rho}_{AA'} = i \bra{A^{(N)}} \dot{\rho} \ket{A'^{(N')}} = (E_A - E_A')  \bra{A^{(N)}}\rho \ket{A'^{(N')}}.
\end{align*}
One can express an eigenstate of the total system in terms of the two subsystems via:
$$\ket{A^{(N)}} = \sum_{a\alpha} p_{a\alpha} \ket{a^{(N_S)}} \otimes \ket{\alpha^{(N_B)}} , $$
with the conservation law $N = N_S(a) + N_B(\alpha)$. 
The full density matrix element thus reads:
$$ \rho_{AA'} = \sum_{aa',\alpha \alpha'} p_{a\alpha}p_{a'\alpha'} \ket{a^{(N_S)}}\bra{a'^{(N_S')}}
\otimes \ket{\alpha^{(N_B)}}  \bra{\alpha'^{(N_B')}} .$$
  Since the reduced density matrix is obtained by a partial trace it follows that $N_B(\alpha) = N_B'(\alpha')$ leading to $N_S'(a') = N_S(a)+ (N'-N )$. Thus the separation of the dynamics with respect to the particle number is transferred to the reduced density matrix and the master equation.
  
  The same observations can be made directly from the master equations~(\ref{equ:integro_differential_equation}) or from the generator, where the question arises which symmetries prevent the occurrence of coherences~\cite{Styliaris2020symmetriesmonotones, cattaneo2020symmetry,Diaz2020accessiblecoherence}.
 \section{\label{app:choi} Choi theorem for the preservation of hermiticity}
Calculating the Choi matrix of a generator $\mathcal{K}$ in the present vectorized notation leads to 
\begin{align*}C(\mathcal{K})=
E_{12} \otimes \mathcal{K}(E_{12})& =   \begin{pmatrix} 0 & 1 \\ 0 & 0 \end{pmatrix} \otimes \mathcal{K} \begin{pmatrix} 0 & 1 \\ 0 & 0 \end{pmatrix} = \begin{pmatrix} 0 & 0 & \mathcal{K}_{11,12} & \mathcal{K}_{12,12} \\ 0 & 0 & \mathcal{K}_{21,12} & \mathcal{K}_{22,12} \\ 0& 0 & 0 & 0\\ 0 & 0 & 0 & 0 \end{pmatrix}.
\end{align*}
Comparing the original operator $\mathcal{K}$ and the Choi matrix $C(\mathcal{K})$ in this column--wise vectorized basis

$$\underbrace{(1,1)}_{1}, \underbrace{(2,1)}_{2}, \dots, \underbrace{(n,1)}_{n}, \underbrace{(1,2)}_{n+1}, \dots (1,n), (2,n), \dots, \underbrace{(n,n)}_{n^2},$$
\begin{align*}
\mathcal{K} & = \begin{pmatrix} \mathcal{K}_{11,11} & \mathcal{K}_{11,21}& \mathcal{K}_{11,12} & \mathcal{K}_{11,22} \\  \mathcal{K}_{21,11} & \mathcal{K}_{21,21}& \mathcal{K}_{21,12} & \mathcal{K}_{21,22} \\\mathcal{K}_{12,11} & \mathcal{K}_{12,21}& \mathcal{K}_{12,12} & \mathcal{K}_{12,22}\\\mathcal{K}_{22,11} & \mathcal{K}_{22,21}& \mathcal{K}_{22,12} & \mathcal{K}_{22,22} \end{pmatrix},\\
C(\mathcal{K}) & = \begin{pmatrix} \mathcal{K}_{11,11} & \mathcal{K}_{12,11} & \mathcal{K}_{11,12} & \mathcal{K}_{12,12} \\ \mathcal{K}_{21,11} & \mathcal{K}_{22,11} & \mathcal{K}_{21,12} & \mathcal{K}_{22,12} \\\mathcal{K}_{11,21} & \mathcal{K}_{12,21} & \mathcal{K}_{11,22} & \mathcal{K}_{12,22} \\ \mathcal{K}_{21,21} & \mathcal{K}_{22,21} & \mathcal{K}_{21,22} & \mathcal{K}_{22,22} \end{pmatrix},
\end{align*} reveals an easy way to calculate $C(\mathcal{K})$ by swapping two indices in $\mathcal{K}$: $$ C(\mathcal{K})_{ij,lm} = \mathcal{K}_{il,jm}.$$
Using this relation we can express~(\ref{equ:K_hermitian}) via the Choi matrices $C$:
\begin{align}
C(\mathcal{K})_{il,jm} \overset{!}{=} C(\mathcal{K})_{jm,il}^*,\quad \Rightarrow \quad C(\mathcal{K})\overset{!}{=} C(\mathcal{K})^\dag,
\end{align} which proves the stated Choi criterion for $\mathcal{K}$ to preserve hermiticity.

\section{\label{sec:appendix_sec_derivation} Microscopic derivation of Born--Markov master equations}
\subsection{Illustration of different types of master equations}
\begin{figure}[ht!]
	\begin{center}
		\includegraphics[width = 1\textwidth]{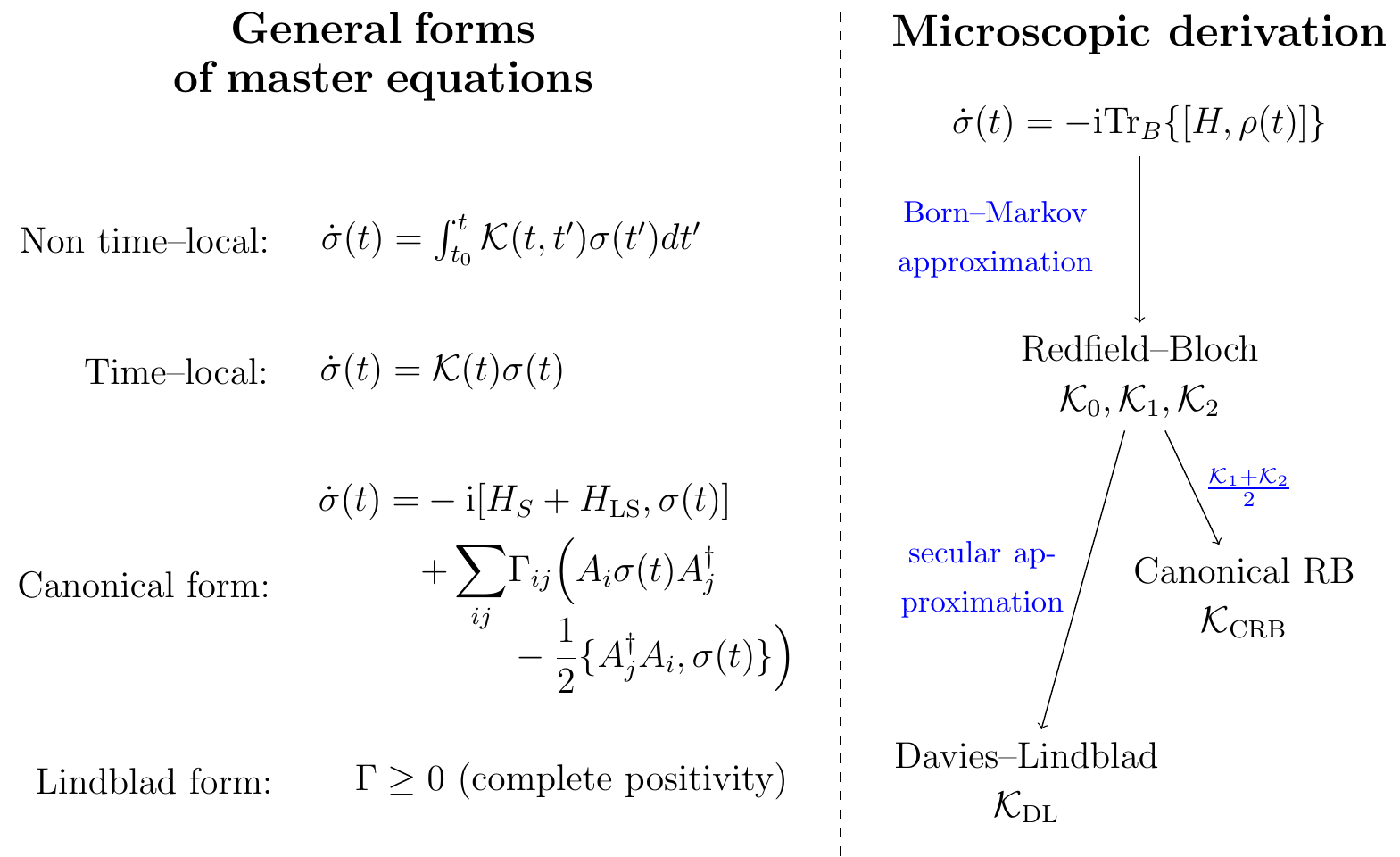}
		\caption{Overview of different forms of quantum master equations in relation with various types of microscopically derived master equations.} \label{fig:master_equations}
	\end{center}
\end{figure}

Figure~\ref{fig:master_equations} gives an overview over types of master equations and the discussed microscopic derivations. Note, that ERMEA can be applied to all discussed microscopically derived master equations.

\subsection{Derivation of the integro--differential equation}\label{app:derivation_BM}
We start with the von--Neumann equation for the full system ($\hbar$ is set to one):
 $$ \dot{\rho}(t) = -\rmi[H, \rho(t)].$$
 We switch to the interaction picture, which will be denoted by bold symbols,\\ $\boldsymbol{\rho}(t) = \rme^{\rmi H_0 (t-t_0)} \rho(t) \rme^{-\rmi H_0 (t-t_0)}$.
 The Schr\"odinger equation in the interaction picture reads:
  $$ \dot{\boldsymbol{\rho}}(t) = -\rmi [\boldsymbol{H}_I(t), \boldsymbol{\rho}(t)].$$
  We integrate this equation from a starting point $t_0$ where system and bath are decoupled $\rho(t_0) = \sigma(t_0) \otimes \rho_b(t_0)$
 $$\boldsymbol{\rho}(t) = \boldsymbol{\rho}(t_0) - \rmi \int_{t_0}^t [\boldsymbol{H}_I(t'), \boldsymbol{\rho}(t')] \rmd t'. $$
 Reinsertion into the Schr\"odinger equation generates a integro--differential equation:
 $$ \dot{\boldsymbol{\rho}}(t) = -\rmi [\boldsymbol{H}_I(t), \boldsymbol{\rho}(t_0)] - \int_0^t \left[\boldsymbol{H}_I(t), [\boldsymbol{H}_I(t'), \boldsymbol{\rho}(t')]\right] \rmd t'.$$
 Going back to the Schr\"odinger picture by using
\begin{align*}
    \frac{\rmd}{\rmd t} \boldsymbol{\rho}(t) = & \rme^{\rmi H_0 (t-t_0)} \rmi [H_0, \rho(t)] \rme^{-\rmi H_0 (t-t_0)} +  \rme^{\rmi H_0(t-t_0)} \left( \frac{\rmd}{\rmd t }\rho(t)\right) \rme^{-\rmi H_0(t-t_0)} ,
\end{align*} and multiplying with $\rme^{-\rmi H_0\underline{t}}$ from the left and $\rme^{\rmi H_0\underline{t}}$ from the right using $\underline{t} = t-t_0$ leads to
 
 \begin{align*}
     \dot{\rho}(t) =& -\rmi[H_0, \rho(t)] - \rmi \rme^{-\rmi H_0\underline{t}} [\rme^{\rmi H_0\underline{t}}H_I \rme^{-\rmi H_0\underline{t}}, \rho(t_0)]\rme^{\rmi H_0\underline{t}}  \\
     &- \int_{t_0}^t \left[H_I, \rme^{-\rmi H_0(t-t')} [H_I, \rho(t')] \rme^{\rmi H_0(t-t')} \right] \rmd t'.
 \end{align*}
 Since we are only interested in a dynamical map of the central system we trace out the bath and define the reduced density operator $\sigma(t) := \Trb{\rho(t)}$. 
 \begin{align*}
     \dot{\sigma}(t) = \,\Trb{\dot{\rho}(t)}
     =& -\rmi\Trb{[H_0, \rho(t)]} - \rmi  \underbrace{\Trb{ [ H_I, \rho(t_0)]}}_{=0}  \\
     & - \int_{t_0}^t \textnormal{Tr}_B \Bigg\{\Big[H_I,  \rme^{-\rmi H_0(t-t')} [H_I, \rho(t')] \cdot \rme^{\rmi H_0(t-t')} \Big]\Bigg\} \rmd t'. 
 \end{align*}The second term vanishes as the interaction Hamiltonian commutes with the decoupled state $\rho(t_0)$.
 The variable transformation ($\tau = t-t'$) with $\rmd t' = -\rmd\tau$,  $\tau \in [t-t_0,0]$ and the flipped limits induce
\begin{align*}
    \dot{\sigma}(t)   =& -\rmi\Trb{[H_0, \rho(t)]} -\int_0^{t-t_0} \rmd\tau\Trb{\left[H_I , \rme^{-\rmi H_0\tau} [H_I, \rho(t-\tau)] \rme^{\rmi H_0\tau}\right]} .
\end{align*}
That way we end up with the still exact non--time--local integro differential equation.
The next step includes sending $t_0$ to minus infinity which can be interpreted as switching on the interaction adiabatically at that time or in the case of a fixed finite $t_0$ by introducing a small error which can be quantified (see~\cite[Equation~(47b)]{mozgunov2020completely}).
The representation of this equation using Liouville operators is given in~(\ref{equ:integro_differential_equation}).

 \subsection{Liouville superoperators} \label{app:superoperators}
 The used Liouville superoperators $\mathcal{L}_I$, $\mathcal{L}_0$ and $\mathcal{L}_S$ are formed by the commutator with the corresponding Hamiltonian $H_I$, $H_0$ and $H_S$, e.g. $\mathcal{L}_I\{A\} = [H_I, A]$, see also~\cite{zwanzig1964identity}. Note, that the exponential of a superoperator formed by a commutator yields $\rme^{-\rmi \mathcal{L}_0 t} \{A\} = \rme^{-\rmi H_0 t} A \rme^{\rmi H_0 t}$. 
 
In the notation used, the superoperators or functions of superoperators act on all subsequent operators unless the effect is indicated by curly brackets as in~(\ref{equ:born_markov_master_equation}). 

Using the eigenbasis $H_S | a\rangle = E_a |a\rangle$ we gain for the evaluation of a function $f$ of the superoperator $\mathcal{L}_S$ acting on the operator $A$ the following important rule 
\begin{equation}
\langle a| f(\mathcal{L}_S) \{A\} | b\rangle = f\left(E_a - E_b\right) \langle a| A | b\rangle. \label{equ:Liouville_functional}
\end{equation}
  \subsection{\label{sec:app_sub_born_approx}Applying Born approximation, derivation of Born--Markov master equation}
 After applying the different types of Markov approximation we perform the Born approximation in the integral terms $\mathcal{D}_0, \mathcal{D}_1$ and $\mathcal{D}_2$ which assumes that the density operator can be written as a tensor product at any time $\rho(t) \approx \rho_B \otimes \sigma(t)$. Inserting the chosen coupling Hamiltonian $H_I$ [see (\ref{equ:coupling})] in the commutators and separating bath and system components lead for $\mathcal{D}_1$ to
 
 \begin{align*}
 \mathcal{D}_1 \approx & -\int_0^\infty \Trb{\Big[ H_I, \rme^{-\rmi H_0 \tau} [H_I, \rho_B \rme^{\rmi H_S \tau}\sigma(t)\rme^{-\rmi H_S\tau} ] \rme^{\rmi H_0 \tau} \Big]}\rmd\tau  \\  
 =&- \int_0^\infty \sum_{\mu\kappa s} \Big[ c_\mu^s,  C_{\mu \kappa}^s(\tau)\rme^{-\rmi H_S\tau} c_{\kappa}^{\overline{s}} \rme^{\rmi H_S\tau} \sigma(t)\Big]\rmd \tau + h.c., 
 \end{align*}
 with the bath correlation function 
 \begin{align}
 C_{\mu \kappa}^s(\tau) &= \sum_{\alpha } V_{\alpha  \mu}^s \langle d_{\alpha}^{\overline{s}}(\tau) d_{\alpha}^s \rangle V_{\alpha  \kappa}^{\overline{s}}. \label{equ:bath_corr_func}
 \end{align} 
 Finally one arrives at the following Born--Markov master equations in the weak coupling limit:

 \tboxit{Redfield--Bloch master equation type $\mathcal{K}_0, \mathcal{K}_1$, $\mathcal{K}_2$ and $\mathcal{K}_{\textnormal{CRB}}$}
 {
 	\begin{subequations} \label{equ:born_markov_master_equation}
 		\begin{align}
 		\mathcal{K}_0\, \sigma(t) =& -\rmi[H_S,\sigma(t)] - \sum_{s\mu\kappa}  \left[c_\mu^s, F_{\mu\kappa}^{s}(-\mathcal{L}_S)\{c_\kappa^{\overline{s}} \sigma (t)\}\right] + \textnormal{h.c.} , \\
 		\mathcal{K}_1 \,\sigma(t)=& -\rmi[H_S,\sigma(t)] - \sum_{s\mu\kappa}  \left[c_\mu^s,  F_{\mu\kappa}^{s}(-\mathcal{L}_S)\{c_\kappa^{\overline{s}}\} \sigma (t)\right] + \textnormal{h.c.}, \\
 		\mathcal{K}_2 \,\sigma(t)=& -\rmi[H_S,\sigma(t)] - \sum_{s\mu\kappa}  \left[F_{\mu\kappa}^{s}(\mathcal{L}_S)\{c_\mu^s\}, c_\kappa^{\overline{s}} \sigma (t)\right] + \textnormal{h.c.}, \\
 		\mathcal{K}_{\textnormal{CRB}} \, \sigma(t)= &{} \frac{\mathcal{K}_1+ \mathcal{K}_2}{2} 	 \,\sigma(t),
 		\end{align}
 	\end{subequations}
 	with the bath contribution term
 	\begin{align}
 	F_{\mu \kappa}^s (\mathcal{L}_S) := \int_0^{\infty} \rmd\tau C_{\mu\kappa}^s(\tau) \rme^{\rmi\mathcal{L_S} \tau}, \label{equ:dissipators}
 	\end{align}
 	and the Liouville superoperator $\mathcal{L_S}$ as defined in~\ref{app:superoperators}.
 }
 \subsection{\label{sec:appendix_decomposition} Decomposition of the bath contribution terms and relation to equilibrium Green functions}
In order to decompose the bath contribution terms [see~(\ref{equ:dissipators})] it is convenient to introduce the Hermitian and anti--Hermitian functions:
\begin{align}
\gamma_{\mu\kappa}^s(\omega) &:= \int_{-\infty}^\infty C_{\mu\kappa}^{s} (\tau) \rme^{\rmi\omega\tau} \rmd\tau,  &&\left(\gamma_{\mu\kappa}^s(\omega) \right)^* =  \gamma_{\kappa\mu}^s(\omega) ,\label{equ:gamma} \\
\sigma_{\mu\kappa}^s(\omega) &:= \int_{-\infty}^\infty C_{\mu\kappa}^{s} (\tau) \textnormal{sign}(\tau) \rme^{\rmi\omega\tau} \rmd\tau,  && \left(\sigma_{\mu\kappa}^s(\omega) \right)^* = - \sigma_{\kappa\mu}^s(\omega),\label{equ:sigma}\\
F_{\mu\kappa}^s(\omega) &= \frac{ \gamma_{\mu\kappa}^s(\omega) +  \sigma_{\mu\kappa}^s(\omega) }{2}. \notag
\end{align}

The aim will be to express the influence of the baths in the equations above in terms of retarded and advanced Green functions $G^R$ and $G^A$.

Starting from~(\ref{equ:bath_corr_func}) we identify the bath correlation functions $C_{\mu\kappa}^s(\tau)$ with the greater and lesser Green functions $G^+ = G^>, G^- = G^<$:
\begin{align}
\langle d_{\alpha}^{\overline{s}}(\tau) d_{\alpha}^s \rangle = \rmi s G^s(s\tau).
\end{align}
The evaluation of this expression in the integral of~(\ref{equ:gamma}) reproduces the Fourier transform of the Green functions.
Since the bath is in equilibrium, greater and lesser Green functions can be expressed via
\begin{align}
G^s(\omega) = s(G^R(\omega) - G^A(\omega)) f^{\overline{s}}(\omega|\beta_\alpha, \mu_\alpha), \notag
\end{align}with the generalized Fermi function $f^+(\omega|\beta_\alpha, \mu_\alpha) = \frac{1}{\exp(\beta_\alpha(\omega - \mu_\alpha)) + 1}$, $f^- = 1-f^+$, the bath temperature $\beta_\alpha$ and the chemical potential $\mu_\alpha$.

Thus the Hermitian part of the bath contribution term becomes:
\begin{align}
\gamma_{\mu\kappa}^s(\omega) = & \sum_{\alpha} V_{\alpha \mu}^s\, \left[ \rmi(G^R_{\alpha}(s\omega) - G^A_{ \alpha}(s\omega)) \cdot f^{\overline{s}}(s\omega|\beta_\alpha, \mu_\alpha) \right]\,V_{\alpha  \kappa}^{\overline{s}}. \label{equ:gamma_Green}
\end{align}
We identified the part in square brackets as bath occupation $O_{\alpha}^s(\omega)$ as defined in~(\ref{equ:occupation}).

For the anti--Hermitian part (\ref{equ:sigma}), the integration can be interpreted as principle value integral:
\begin{align*}
\int_{-\infty}^\infty g(\tau) \textnormal{sign}(\tau) \rme^{\rmi\omega \tau} \rmd\tau &= \int_{-\infty}^\infty \frac{\rmd\omega'}{2\pi} g(\omega') \int_{-\infty}^\infty \textnormal{sign}(t) \rme^{\rmi(\omega - \omega')\tau - \varepsilon |\tau|} \rmd\tau \\
&= \frac{\rmi}{\pi} \int_{-\infty}^\infty g(\omega') \frac{\overbrace{\omega - \omega'}^{x:=}}{(\omega-\omega')^2 + \varepsilon^2} \rmd\omega'  = \frac{\rmi}{\pi} \mathcal{P} \int_{-\infty}^\infty \frac{g(\omega-x)}{x} \rmd x,
\end{align*}leading to~(\ref{equ:sigma_full}).

\section{Secular approximation and Davies--Lindblad master equation \label{sec:appendix_secBM_lindblad}}
Starting from Redfield--Bloch type master equations $\mathcal{K}_1$ [see~(\ref{equ:born_markov_master_equation})] and after decomposing the dissipators $F$ as illustrated in~\ref{sec:appendix_decomposition} we gain 
\begin{align}
\mathcal{K}_1\, \sigma(t) = -\rmi [H_S, \sigma(t)] -\frac{1}{2}\sum_{s\mu\kappa} & \left[ c_\mu^s,\, \gamma_{\mu\kappa}^{s}(-\mathcal{L})\{c_\kappa^{\overline{s}}\} \sigma (t)\right] + \left[ c_\mu^s,\,\sigma_{\mu \kappa}^{s}(-\mathcal{L})\{c_\kappa^{\overline{s}}\} \sigma (t)\right] \notag \\
& + \left[ \sigma(t) \gamma_{\mu\kappa}^{s}(\mathcal{L})\{c_{\mu}^s\}, \, c_\kappa^{\overline{s}}\right] - \left[ \sigma(t)  \sigma_{\mu\kappa}^{s}(\mathcal{L})\{c_{\mu}^s\}, \, c_\kappa^{\overline{s}}\right]. \label{equ:born_markov_expanded}
\end{align}
Inserting full sets of eigenbasis vectors of the system Hamiltonian, having $H_S \ket{a} = E_a \ket{a}$, with the eigenvectors $\ket{a}$ and eigenenergies $E_a$, induces in the first term of the first commutator
\begin{align}
\sum_{abcde}\ket{a}\bra{a} c_\mu^s \ket{b} \bra{b}\ket{d}\bra{d} \gamma_{\mu\kappa}^s(-\mathcal{L})\{c_\kappa^{\overline{s}} \}  \ket{c} \bra{c} \sigma(t)\ket{e}\bra{e}. \label{equ:eigenstate_notation}
\end{align}The set of eigenvectors $\ket{d}$ was inserted to motivate the later used traceless operators $A_i$ in the Lindblad equation. It holds $E_b = E_d$. 

The \textbf{secular approximation} is applied via a projection superoperator $\mathcal{P}_E$ that leaves a reduced density matrix block diagonal in its eigenenergies:
$$\sigma_E := \mathcal{P}_E\, \sigma=  \mathcal{P}_E \sum_{ab} \sigma_{ab} \ket{a} \bra{b} = \sum_{ab} \sigma_{ab} \ket{a}\bra{b} \delta_{E_aE_b}.$$
Applying this projection on the master equation and the reduced density operator $$\mathcal{P}_E \, \dot{\sigma} = \mathcal{P}_E \,\mathcal{K} \mathcal{P}_E\,\sigma = :\mathcal{K}_{\textnormal{DL}} \sigma_E,$$
implies  $E_a = E_c = E_e$ in (\ref{equ:eigenstate_notation}). 

Since functions of the Liouville superoperator $\mathcal{L}_S$ are evaluated according to~(\ref{equ:Liouville_functional}) and only two energies $E_a$ and $E_b$ occur in the secular approximation, we can always switch the action of the bath contribution terms on the creation and annihilation operators: 
\begin{align}
\bra{a} c_\mu^s \ket{b} \bra{b} \gamma^s_{\mu\kappa} (-(E_b - E_a)) c_\kappa^{\overline{s}} \ket{c}\delta_{E_aE_c}  = \gamma^s_{\mu\kappa}(E_a - E_b) \bra{a} c_\mu^s \ket{b} \bra{b} c_\kappa^{\overline{s}} \ket{c}\delta_{E_aE_c},
\end{align} So in the secular approximation the following relation holds
\begin{align}
\mathcal{P}_E\,  c_\mu^s F_{\mu\kappa}^s(-\mathcal{L}_S)\{c_\kappa^{\overline{s}}\sigma_E(t)\} = \mathcal{P}_E c_\mu^s  F_{\mu\kappa}^s(-\mathcal{L}_S)\{c_\kappa^{\overline{s}} \} \sigma_E(t) = \mathcal{P}_E\,F_{\mu\kappa}^s(\mathcal{L}_S) \{c_\mu^s\} c_\kappa^{\overline{s}} \sigma_E(t), \label{equ:sec_swapping}
\end{align}
 that proofs equivalence of all introduced Redfield--Bloch master equations~(\ref{equ:born_markov_master_equation}) in the secular limit.
 
 Finally we can condense (\ref{equ:born_markov_expanded}) 
\begin{align}
\mathcal{K}_{\textnormal{DL}}\sigma_E= -\rmi [H_S, \sigma_E(t)] + \mathcal{P}_E \sum_{s\mu\kappa}& c_\kappa^{\overline{s}} \sigma_E(t) \gamma_{\mu\kappa}^s(\mathcal{L})\{c_\mu^s\} -\frac{1}{2} \{ \gamma_{\mu\kappa}^s(\mathcal{L})\{c_\mu^s\} c_\kappa^{\overline{s}}, \sigma_E(t)\}\notag \\
&+ [\sigma_{\mu\kappa}^s(\mathcal{L})\{c_\mu^s\} c_\kappa^{\overline{s}}, \sigma_E(t)], \label{equ:decomposed_K1}
\end{align}
to gain the so--called Davies--Lindblad master equation $\mathcal{K}_{\textnormal{DL}}$ in Lindblad form~(\ref{equ:lindblad_equation}):

\tboxit{Davies--Lindblad master equation}{
	\begin{align}
\mathcal{K}_{\textnormal{DL}}\sigma(t) = & -\rmi[H_S + H_{\textnormal{LS}}, \sigma(t)] + \sum_{ij} \Gamma_{ij}^{\textnormal{DL}}\Big(A_i \sigma(t) A_j^\dag  - \frac{1}{2} \{A_j^\dag A_i, \sigma(t)\}\Big), \label{equ:davies_lindblad}
\end{align}
using the traceless operators $A_i = \ket{a}\bra{b}$  and $A_j = \ket{c}\bra{d}$ with the collective indices $i=(a,b)$, $j=(c,d)$ and system eigenvectors $\ket{a},\ket{b},\ket{c},\ket{d}$. 

The Gamma matrix $\Gamma$ and the Lamb--shift Hamiltonian $H_{\textnormal{LS}}$ are given by:
\begin{align}
\Gamma_{ij}^{\textnormal{DL}}& = \sum_{\mu\kappa}  \bra{a} c_\kappa^{\overline{s}} \ket{b}\cdot \gamma_{\mu\kappa}^s(E_{a} - E_b) \cdot  \bra{d} c_\mu^s \ket{c} \delta_{E_aE_c} \delta_{E_bE_d},\label{equ:gamma_davies_lindblad}\\
H_{\textnormal{LS}} & = -\rmi\mathcal{P}_E\sum_{s\mu\kappa} \sigma_{\mu\kappa}^s(\mathcal{L})\{c_\mu^s\} c_\kappa^{\overline{s}}.  \notag
\end{align}
Note that $s$ is determined by the choice of $i$ since the state $b$ is in the particle sector $N+s$ given that $a$ is in sector $N$. For a nonzero entry in $\Gamma$ index $j$ has to follow this diction of particle sectors since to particle conservation.
}
The Gamma matrix in the Davies--Lindblad approach is a submatrix of the Gamma matrices used in the other Redfield--Bloch master equations.

 \section{\label{sec:app_current} Current formula for a Lindblad type master equation}
Referring to the Lindblad equation [see~(\ref{equ:lindblad_equation})], the current is given by the time derivative of the total number of particles $\hat{N}$
 \begin{align*}
     I(t) =& \frac{e}{\hbar} \frac{\rmd}{\rmd t}\langle \hat{N}(t) \rangle = \frac{e}{\hbar} \frac{\rmd}{\rmd t} \Tr{\hat{N} \sigma(t)} \\
     =& -\rmi\frac{e}{\hbar^2} \underbrace{\Tr{ \hat{N} [H_S + H_{\textnormal{LS}}, \sigma(t)]}}_{\Tr{[\hat{N}, H_S+H_{\textnormal{LS}}] \sigma(t)} = 0}  + \frac{e}{\hbar} \Tr{ \frac{1}{2} \sum_{ij} \Gamma_{ij}\hat{N} [A_i,\sigma(t) A_j^\dag] + h.c. } \\
     =&\frac{e}{\hbar} \sum_{ij}\textnormal{Re} \Tr{ \Gamma_{ij} [\hat{N},A_i] \sigma(t)A_j^\dag}.
 \end{align*}
 Using the eigenbasis operators and with the definition of $s(a,b)$ in~(\ref{equ:Gamma_vector}) we obtain for the current along bath $\alpha$ 
 \begin{align}
I_\alpha(t)&= \frac{e}{\hbar} \sum_{ab,cd} \textnormal{Re}\Tr{\Gamma_{ab|cd}^\alpha \underbrace{[\hat{N},\ket{a}\bra{b}]}_{=N_a - N_b = -s} \sigma \ket{d} \bra{c}  } =- \frac{e}{\hbar} \sum_{abd} s(a,b)\textnormal{Re}\Gamma_{ab|ad}^\alpha \bra{b} \sigma \ket{d}.
\end{align}

\section{Discussion of errors and comparison with exact results}
\label{app:discussion_of_errors}
\subsection{Validity of Born--Markov master equations and comparison with exact results}
In order to illustrate the applicability of the discussed Born--Markov master equations we compare CRB master equations results with exact results derived via non--equilibrium Green function (NEGF) techniques in the non--interacting case.

The examined spinless fermionic three--site model is described by the Hamiltonian
$$ H =  \sum_{ij=1}^3  T_{ij} c_i^\dag c_j + V \sum_{i<j} n_i n_j,$$
with $V = 0$ eV (non--interacting case) and $$T = \begin{pmatrix}
	1 & -0.1 & 0 \\ -0.1 & 1 & -0.1 \\ 0 & -0.1 & 6
\end{pmatrix}, $$ given in units of electronvolt. The same bath setup is used as in Section~\ref{sec:num_res_ERMEA} with the coupling Hamiltonian given by (\ref{equ:coupling}) and with a variable coupling strength $V_{L,1} = V_{R,2} = V$.

The exact current from the left lead into the non--interacting quantum system is given by~\cite{meir1992landauer, ness2010generalization}:
$$I_L = \frac{e}{h} \int_{-\infty}^\infty \textnormal{Re}\left[ \textnormal{Tr}_S \left\{\boldsymbol{V}_{SB_L} \cdot \boldsymbol{G}_{B_LS}^K(\omega) \right\}\right] d\omega,$$ with the system--bath coupling $\boldsymbol{V}_{SB_L}$ and the Keldysh part $\boldsymbol{G}^K_{B_L S}$ of the full Green function from system to bath which can be exact evaluated using the Dyson--like hybridization equation $\boldsymbol{G}^{-1} = \boldsymbol{g}^{-1} - \Delta$, that needs the isolated Green function $\boldsymbol{g}$ and the hybridization function $\Delta = \sum_\alpha \boldsymbol{V}_{SB_\alpha} \boldsymbol{g}_\alpha \boldsymbol{V}_{B_\alpha S}$ with the equilibrium bath Green functions $\boldsymbol{g}_\alpha$ of left and right lead.

The current formula is equivalent to the Landauer--Büttiker formula
$$I = \frac{e}{h} \int_{-\infty}^{\infty} d\omega (f_L(\omega|\beta_L, \mu_L) - f_R(\omega|\beta_R, \mu_R)) T(\omega), $$ 
with the transmission function $T(\omega) = \Tr\{\boldsymbol{\Gamma}_L(\omega) \boldsymbol{G}^R(\omega) \boldsymbol{\Gamma}_R(\omega) \boldsymbol{G}^A(\omega)\}$ obtained by the full Green function of the system $\boldsymbol{G}$ and the imaginary part of the hybridization function $\boldsymbol{\Gamma} = \frac{\boldsymbol{\Delta} - \boldsymbol{\Delta}^\dag}{2i}$.
	
\begin{figure}
	\begin{center}
		\includegraphics[width= 0.9\textwidth]{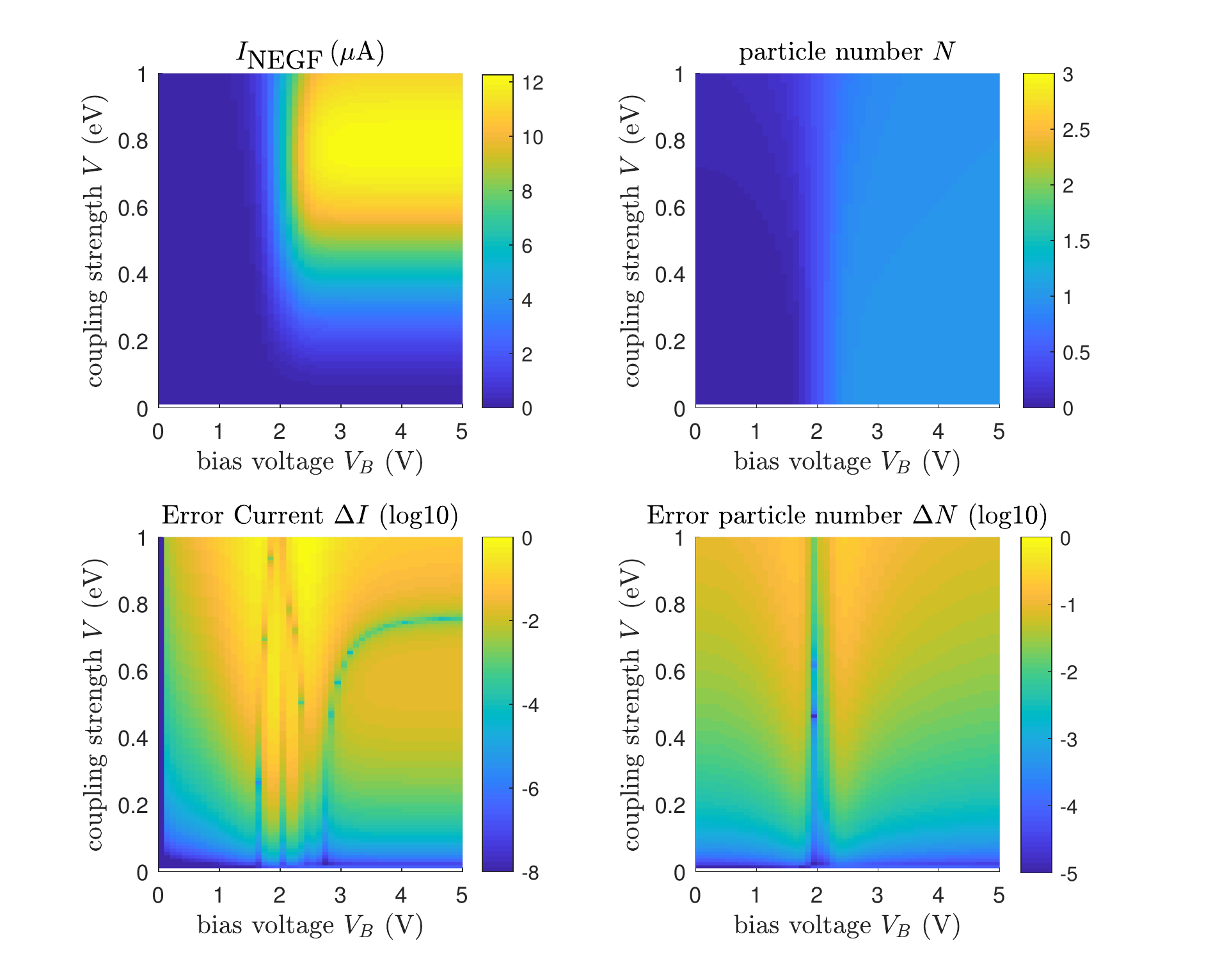}
		\caption{Comparison of Born--Markov master equations with exact NEGF results for a non--interacting spinless three--site model.\\
		The upper left and right panels show the exact current $I_{\textnormal{NEGF}}$ and particle number $N_{\textnormal{NEGF}}$ in the system for different applied bias voltage $V_B$ and coupling strength $V$.\\
	   The two lower panels show the error of the current $\Delta I$ and the error of the particle number $\Delta N$ obtained from the CRB master equation.} \label{fig:exact_results}
	\end{center}
\end{figure}

In Figure~\ref{fig:exact_results} we depict the exact current $I_{\textnormal{NEGF}}$ and particle number $N_{\textnormal{NEGF}}$ derived by the non--equilibrium Green function techniques
and the relative error $\Delta I$ and $\Delta N$ to the current and particle number derived from the full CRB master equation under the variation of the coupling strength $V$ and the applied bias voltage $V_B$. 

Figure~\ref{fig:exact_results} exemplifies the validity of master equations in the weak coupling regime (i.e. $V < 0.1$ eV)  where they coincide with exact results in the non--interacting case.

\subsection{Impact of the coupling strength and bias voltage on the error induced by the partial secular approximation within CRB master equation}
We want to examine the validity of the partial secular approximation for an interacting fermionic spinless three--site model using the same setup as the previous subsection but with an interaction strength $V = -1.7$ eV yielding the spectrum and structure of the density matrix as depicted in Figure~\ref{fig:reduction_of_variables}.

\begin{figure}
	\begin{center}
		\includegraphics[width= 0.8\textwidth]{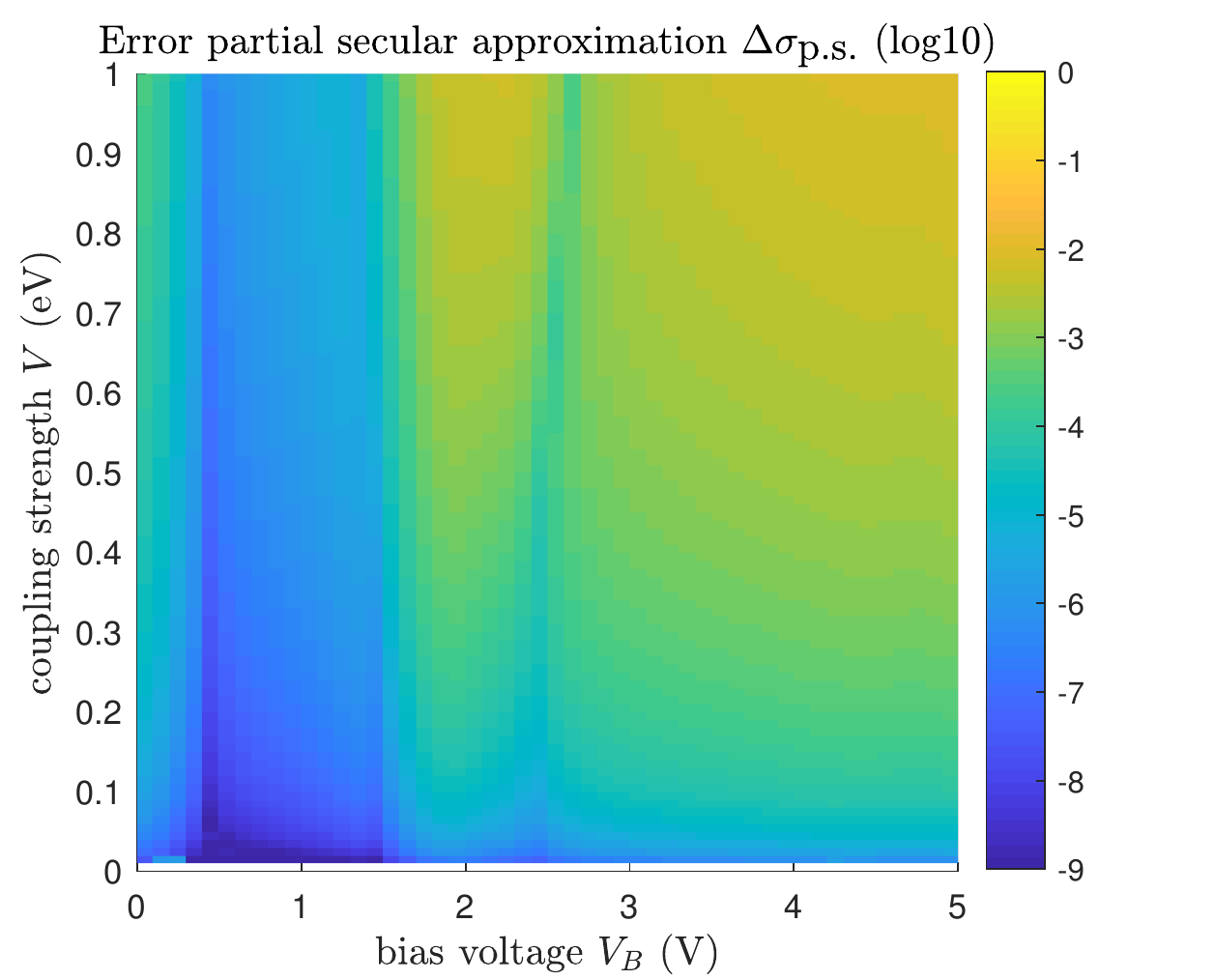}
		\caption{Hilbert--Schmidt error of the reduced density matrices obtained from CRB master equation with and without the partial secular approximation ($\Delta E = 0.2$~eV) for different applied bias voltage $V_B$ and coupling strength $V$. } \label{fig:partial_secular_approximation}
	\end{center}
\end{figure}

Figure~\ref{fig:partial_secular_approximation} illustrates the error of the steady state density matrix induced by the partial secular approximation (using the threshold energy gap $\Delta E = 0.2$ eV) for different coupling strength $V$ and applied bias voltage~$V_B$. 
The error  $\Delta \sigma_{\textnormal{p.s.}}$ is given by the Hilbert--Schmidt norm of the difference of the two reduced density matrices obtained from the CRB master equation with and without the partial secular approximation:
$$\Delta \sigma_{\textnormal{p.s.}} = \lVert \sigma_{\textnormal{p.s.}} - \sigma \rVert_{\textnormal{HS}} = \sqrt{\Tr{(\sigma_{\textnormal{p.s.}} - \sigma)^{2}}}. $$
For Hermitian operators the Hilbert--Schmidt norm is equal to the Frobenius norm.

The partial secular approximation remains valid in the weak--coupling regime (i.e.~$V \lesssim 0.5 \,\Delta E = 0.1$ eV).


\subsection{Spinless six site model, decomposition of the superoperator and error analysis of ERMEA}

To illustrate the effectiveness of the ERMEA approach we discuss a spinless six site model which is derived from the one discussed in Section~\ref{sec:num_res_ERMEA} by restricting the spin in the Hamiltonian (\ref{equ:Hamiltonian}) to $\sigma = \sigma' = \,\uparrow$ and using the same bath and coupling setup.

\begin{figure}
	\begin{center}
		\includegraphics[width=0.8\textwidth]{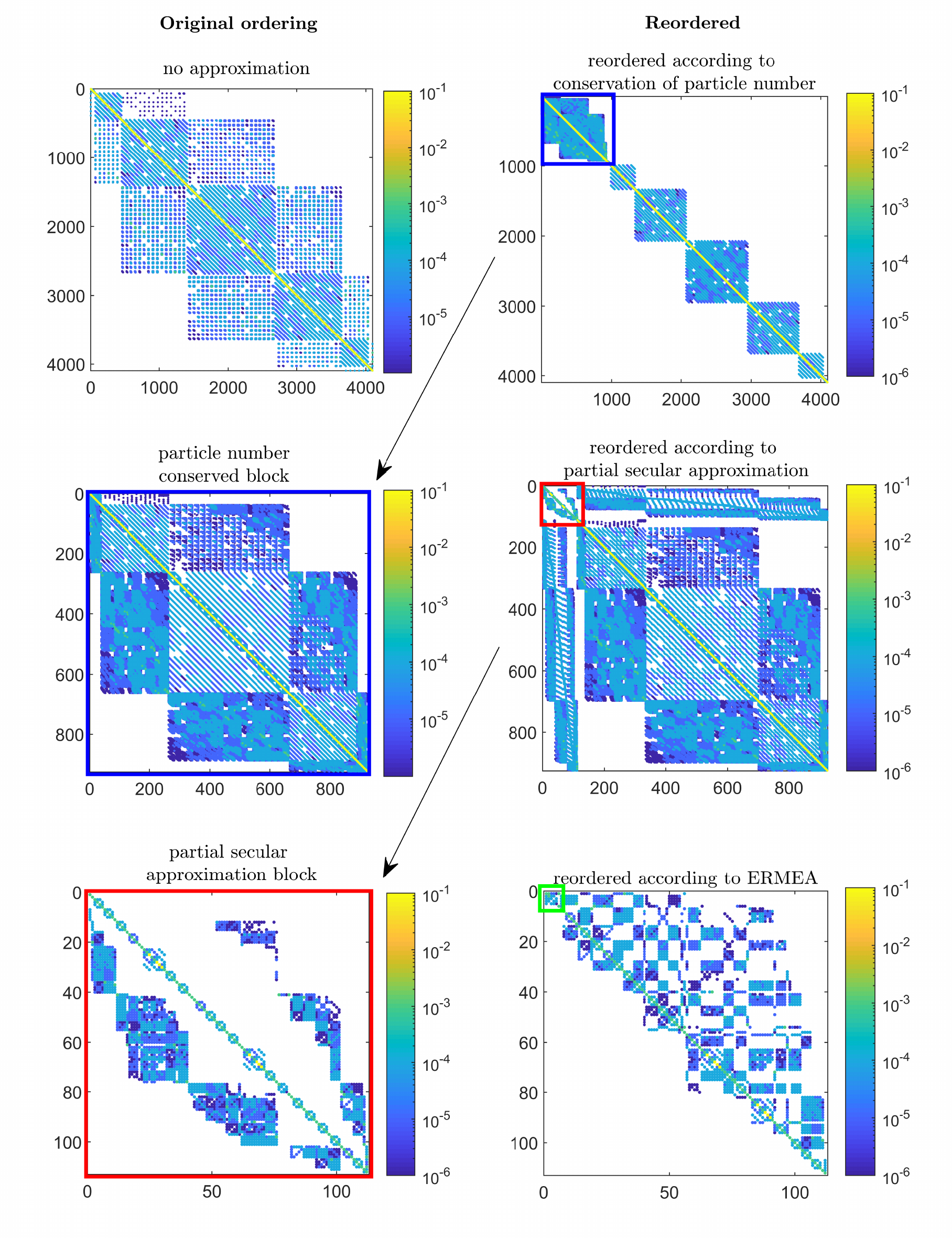}
		\caption{Illustration of the reduction of variables of the superoperator of the CRB master equation for the spinless six site model with a coupling strength $V=0.1$ eV and a bias voltage of $V_B = 2$ V. All six panels show the log10 absolute values of the reduced and reordered superoperator.  \\
		The upper two panels show the full superoperator and a reordering according to entries of the density matrix that preserve the particle number and entries that don't. \\
		The middle two panels show the relevant block (blue) of particle preserved dynamics and a reordering according to the relevant and irrelevant coherences (non--diagonal entries in the density matrix) of a partial secular approximation with $\Delta E = 0.2$ eV. \\
		The lower two panels display the relevant block (red) due to the partial secular approximation and a reshuffling according to the energy cost function used in the ERMEA approach. The green square in the right panel illustrates the relevant part for converged ERMEA.} \label{fig:reordered}
	\end{center}
\end{figure}

The three reductions of variables of the full superoperator $\mathcal{K}^{(\textnormal{CRB})}$ are visualized in Figure~\ref{fig:reordered} which depicts the modulus of the elements of the superoperator for a fixed coupling strength $V = V_{L,1} = V_{R,3} = 0.1$ eV and a bias voltage of $V_B = 2$ V. The right column shows a reordering of the superoperator according to particle number (first row), according to considered and ignored coherences (partial secular approximation in the second row) and according to the average grandcanonical energies as considered in ERMEA (third row).

The error due to the partial secular approximation with $\Delta E = 0.2$ eV amounts to $\Delta \sigma_{\textnormal{p.s.}} = 2.1\cdot10^{-4} $.
The error due to the ERMEA approach with $\nu_c = 10^{-10}$ amounts to $\Delta \sigma_{\mathcal{E}} = 2.2\cdot 10^{-14}$ and the total error is dominated by the error due to the partial secular approximation with $\Delta \sigma_{\textnormal{p.s.},\mathcal{E}} = 2.1\cdot 10^{-4}$.

\begin{figure}
	\begin{center}
		\includegraphics[width=0.8\textwidth]{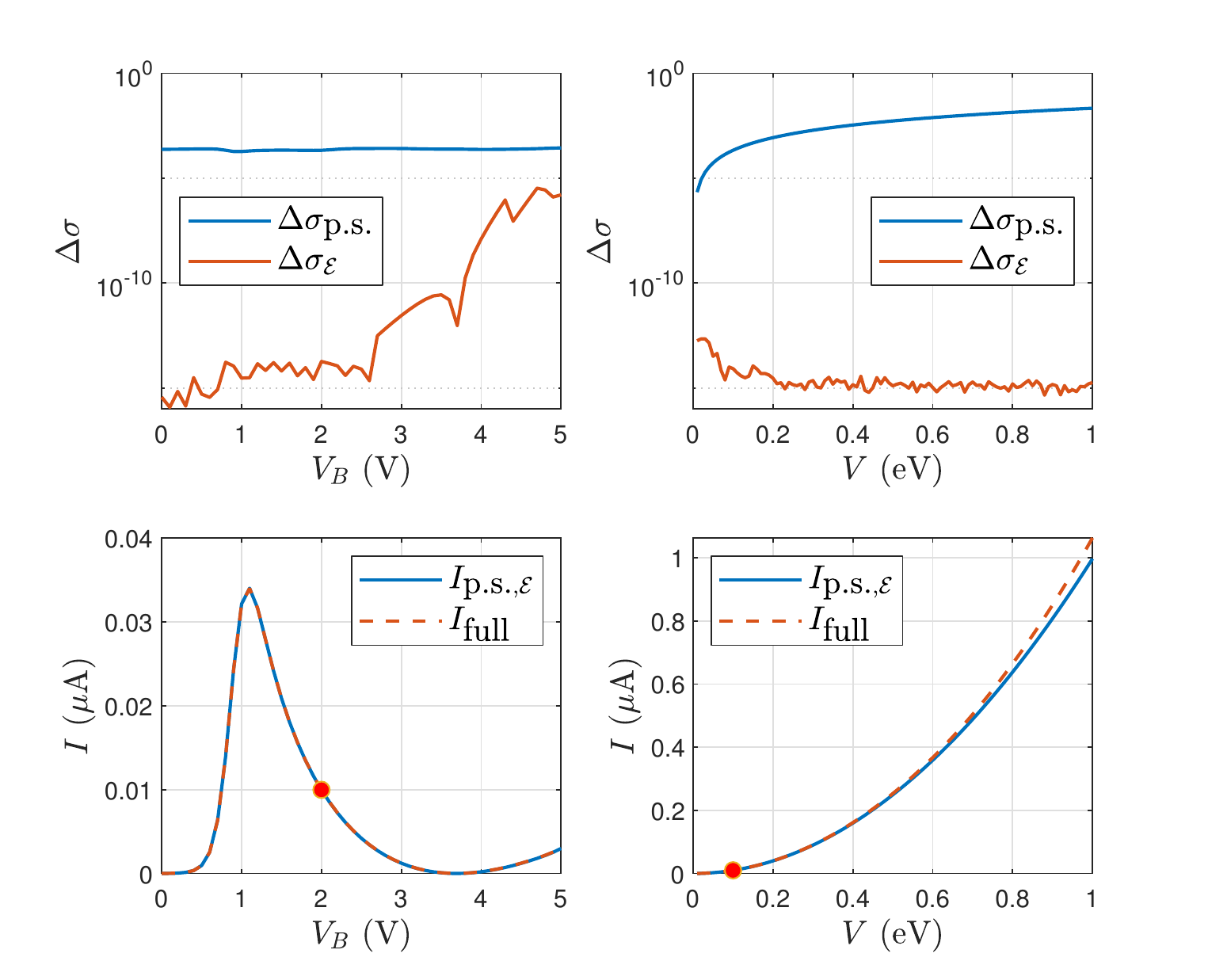}
		\caption{Error analysis of the spinless six--site model of combined partial secular approximation ($\Delta E = 0.2$ eV) and ERMEA ($\nu_c = 10^{-11}$) for a variation of the bias voltage $V_B$ (left panels, fixed $V = 0.1$ eV) and coupling strength $V$ (right panels, fixed $V_B = 2$ V). \\
		The upper panels indicate the error $\Delta \sigma_{\textnormal{p.s.}}$ due to the partial secular approximation and the error $\Delta \sigma_\mathcal{E}$ due to the ERMEA approximation. \\
		The lower panels show the calculated current $I_{\textnormal{full}}$ of the full CRB master equation and the current $I_{\textnormal{p.s.},\mathcal{E}}$ obtained using both approximations.} \label{fig:site_6_errors}
	\end{center}
\end{figure}

Figure~\ref{fig:site_6_errors} illustrates the contributions to the total error due to a combination of partial secular approximation and ERMEA for the variation of the bias voltage $V_B$ (left panel, fixed $V = 0.1$ eV) and coupling strength $V$ (right panel, fixed $V_B = 2$ V). The error is dominated by the partial secular approximation but does not show significant implications in the current characteristics for the weak--coupling limit.

\begin{figure}
	\begin{center}
 		\includegraphics[width=1\textwidth]{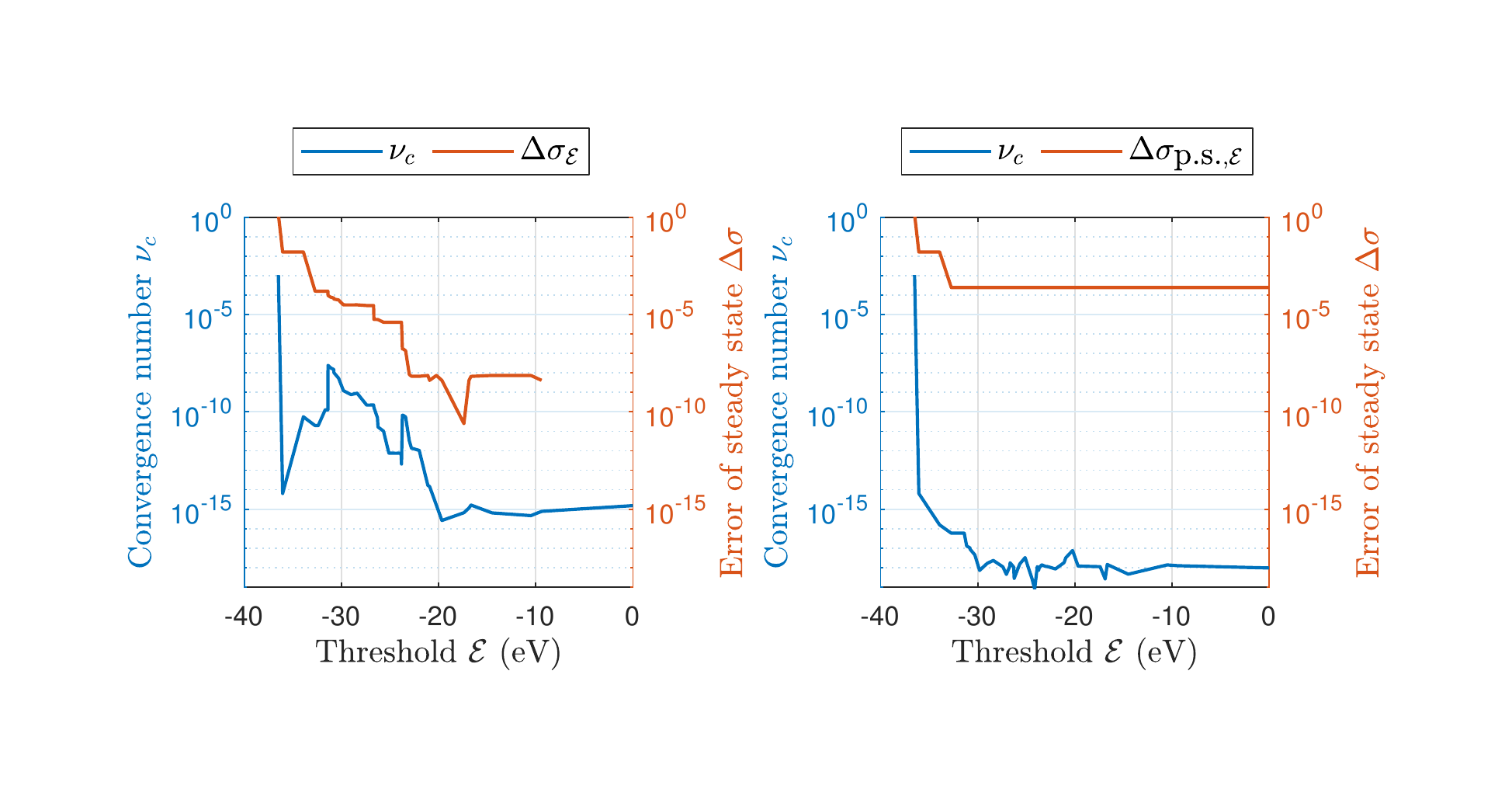}
		\caption{Error of the steady state density matrix $\Delta \sigma$ and convergence number $\nu_c$ of the ERMEA approach without (left panel) and with (right panel) the partial secular approximation ($\Delta E = 0.2$ eV) for a fixed bias voltage $V_B = 3.6$ V and a coupling strength $V = 0.1$ eV.} \label{fig:errors_ERMEA}
	\end{center}
\end{figure}

Finally Figure~\ref{fig:errors_ERMEA} shows the error of ERMEA $\Delta \sigma_{\mathcal{E}}$ and $\Delta \sigma_{\textnormal{p.s.},\mathcal{E}}$ without and with the partial secular approximation compared with the steady--state density matrix of the full CRB master equation. The errors of the density matrices (given by the Frobeniusnorm) and the ERMEA convergence factors $\nu_c$ are depicted as a function of the energy threshold $\mathcal{E}$.

This illustrates the fast continuous convergence of the ERMEA approach as also described in Figure~\ref{fig:convergence_trend_fixed_Vb}.

\section*{References}

 \bibliographystyle{iopart-num}
 \bibliography{literature}{}

\providecommand{\newblock}{}
\begin{thebibliography}{10}
\expandafter\ifx\csname url\endcsname\relax
  \def\url#1{{\tt #1}}\fi
\expandafter\ifx\csname urlprefix\endcsname\relax\def\urlprefix{URL }\fi
\providecommand{\eprint}[2][]{\url{#2}}

\bibitem{papior2017improvements}
Papior N, Lorente N, Frederiksen T, Garc{\'\i}a A and Brandbyge M 2017 {\em
  Computer Physics Communications\/} {\bf 212} 8--24

\bibitem{novaes2006density}
Novaes F~D, da~Silva A~J and Fazzio A 2006 {\em Brazilian Journal of Physics\/}
  {\bf 36} 799--807

\bibitem{gorini1976completely}
Gorini V, Kossakowski A and Sudarshan E~C~G 1976 {\em Journal of Mathematical
  Physics\/} {\bf 17} 821--825

\bibitem{Lindblad1976}
Lindblad G 1976 {\em Communications in Mathematical Physics\/} {\bf 48}
  119--130 ISSN 1432-0916 \urlprefix\url{http://dx.doi.org/10.1007/BF01608499}

\bibitem{breuer2009measure}
Breuer H~P, Laine E~M and Piilo J 2009 {\em Physical review letters\/} {\bf
  103} 210401

\bibitem{whitney2008staying}
Whitney R~S 2008 {\em Journal of Physics A: Mathematical and Theoretical\/}
  {\bf 41} 175304

\bibitem{wenderoth2016sharp}
Wenderoth S, B{\"a}tge J and H{\"a}rtle R 2016 {\em Physical Review B\/} {\bf
  94} 121303

\bibitem{schinabeck2016hierarchical}
Schinabeck C, Erpenbeck A, H{\"a}rtle R and Thoss M 2016 {\em Physical Review
  B\/} {\bf 94} 201407

\bibitem{DordaNussEtAl2014}
Dorda A, Nuss M, von~der Linden W and Arrigoni E 2014 {\em Phys. Rev. B\/} {\bf
  89}(16) 165105
  \urlprefix\url{http://link.aps.org/doi/10.1103/PhysRevB.89.165105}

\bibitem{sorantin2019auxiliary}
Sorantin M~E, Fugger D~M, Dorda A, von~der Linden W and Arrigoni E 2019 {\em
  Physical Review E\/} {\bf 99} 043303

\bibitem{schultz2009quantum}
Schultz M~G and von Oppen F 2009 {\em Physical Review B\/} {\bf 80} 033302

\bibitem{jeske2015bloch}
Jeske J, Ing D~J, Plenio M~B, Huelga S~F and Cole J~H 2015 {\em The Journal of
  chemical physics\/} {\bf 142} 064104

\bibitem{eastham2016bath}
Eastham P, Kirton P, Cammack H, Lovett B and Keeling J 2016 {\em Physical
  Review A\/} {\bf 94} 012110

\bibitem{dominy2016beyond}
Dominy J~M and Lidar D~A 2016 {\em Quantum Information Processing\/} {\bf 15}
  1349--1360

\bibitem{begemann2009quantum}
Begemann G 2009 {\em Quantum interference phenomena in transport through
  molecules and multiple quantum dots\/} vol~9 (Universit{\"a}tsverlag
  Regensburg)

\bibitem{rivas2012open}
Rivas A and Huelga S~F 2012 {\em Open quantum systems\/} (Springer)

\bibitem{kraus1983states}
Kraus K, B{\"o}hm A, Dollard J~D and Wootters W 1983 {\em Lecture notes in
  physics\/} {\bf 190}

\bibitem{hall2014canonical}
Hall M~J, Cresser J~D, Li L and Andersson E 2014 {\em Physical Review A\/} {\bf
  89} 042120

\bibitem{nielsen2002quantum}
Nielsen M~A and Chuang I 2002 Quantum computation and quantum information

\bibitem{rivas2010markovian}
Rivas A, Plato A~D~K, Huelga S~F and Plenio M~B 2010 {\em New Journal of
  Physics\/} {\bf 12} 113032

\bibitem{mitchison2018non}
Mitchison M~T and Plenio M~B 2018 {\em New Journal of Physics\/} {\bf 20}
  033005

\bibitem{barrat1961c}
Barrat J and Cohen-Tannoudji C 1961 {\em J. Phys. Rad\/} {\bf 22} 329

\bibitem{happer1972optical}
Happer W 1972 {\em Reviews of Modern Physics\/} {\bf 44} 169

\bibitem{dumcke1979proper}
D{\"u}mcke R and Spohn H 1979 {\em Zeitschrift f{\"u}r Physik B Condensed
  Matter\/} {\bf 34} 419--422

\bibitem{Farina2019coarse}
Farina D and Giovannetti V 2019 {\em Phys. Rev. A\/} {\bf 100}(1) 012107
  \urlprefix\url{https://link.aps.org/doi/10.1103/PhysRevA.100.012107}

\bibitem{cresser2017coarse}
Cresser J and Facer C 2017 {\em arXiv preprint arXiv:1710.09939\/}

\bibitem{Cattaneo_2019}
Cattaneo M, Giorgi G~L, Maniscalco S and Zambrini R 2019 {\em New Journal of
  Physics\/} {\bf 21} 113045

\bibitem{choi1975completely}
Choi M~D 1975 {\em Linear algebra and its applications\/} {\bf 10} 285--290

\bibitem{havel2003robust}
Havel T~F 2003 {\em Journal of Mathematical Physics\/} {\bf 44} 534--557

\bibitem{bengtsson2017geometry}
Bengtsson I and {\.Z}yczkowski K 2017 {\em Geometry of quantum states: an
  introduction to quantum entanglement\/} (Cambridge university press)

\bibitem{rivas2010entanglement}
Rivas {\'A}, Huelga S~F and Plenio M~B 2010 {\em Physical review letters\/}
  {\bf 105} 050403

\bibitem{pastuszak2020quantifier}
Pastuszak G, Skowyrski A and Jamio{\l}kowski A 2020 {\em arXiv preprint
  arXiv:2003.09081\/}

\bibitem{kossakowski1972quantum}
Kossakowski A 1972 {\em Reports on Mathematical Physics\/} {\bf 3} 247--274

\bibitem{Breuer2002}
Breuer H~P and Petruccione F 2002 {\em The theory of open quantum systems\/}
  (Oxford University Press on Demand)

\bibitem{Schaller2014}
Schaller G 2014 {\em Open Quantum Systems Far from Equilibrium\/} vol 881
  (Springer)

\bibitem{schaller2008preservation}
Schaller G and Brandes T 2008 {\em Physical Review A\/} {\bf 78} 022106

\bibitem{majenz2013coarse}
Majenz C, Albash T, Breuer H~P and Lidar D~A 2013 {\em Physical Review A\/}
  {\bf 88} 012103

\bibitem{mozgunov2020completely}
Mozgunov E and Lidar D 2020 {\em Quantum\/} {\bf 4} 227

\bibitem{kirvsanskas2018phenomenological}
Kir{\v{s}}anskas G, Francki{\'e} M and Wacker A 2018 {\em Physical Review B\/}
  {\bf 97} 035432

\bibitem{potts2019introduction}
Potts P~P 2019 {\em arXiv preprint arXiv:1906.07439\/}

\bibitem{ptaszynski2019thermodynamics}
Ptaszy{\'n}ski K and Esposito M 2019 {\em Physical review letters\/} {\bf 122}
  150603

\bibitem{rumetshofer2017first}
Rumetshofer M, Dorn G, Boeri L, Arrigoni E and von~der Linden W 2017 {\em New
  journal of physics\/} {\bf 19} 103007

\bibitem{DarauBegemannEtAl2009}
Darau D, Begemann G, Donarini A and Grifoni M 2009 {\em Physical Review B\/}
  {\bf 79} 235404 ISSN 1550-235X
  \urlprefix\url{http://dx.doi.org/10.1103/PhysRevB.79.235404}

\bibitem{markussen2010relation}
Markussen T, Stadler R and Thygesen K~S 2010 {\em Nano letters\/} {\bf 10}
  4260--4265

\bibitem{Styliaris2020symmetriesmonotones}
Styliaris G and Zanardi P 2020 {\em {Quantum}\/} {\bf 4} 261 ISSN 2521-327X
  \urlprefix\url{https://doi.org/10.22331/q-2020-04-30-261}

\bibitem{cattaneo2020symmetry}
Cattaneo M, Giorgi G~L, Maniscalco S and Zambrini R 2020 {\em Physical Review
  A\/} {\bf 101} 042108

\bibitem{Diaz2020accessiblecoherence}
D{\'{i}}az M~G, Desef B, Rosati M, Egloff D, Calsamiglia J, Smirne A,
  Skotiniotis M and Huelga S~F 2020 {\em {Quantum}\/} {\bf 4} 249 ISSN
  2521-327X \urlprefix\url{https://doi.org/10.22331/q-2020-04-02-249}

\bibitem{zwanzig1964identity}
Zwanzig R 1964 {\em Physica\/} {\bf 30} 1109--1123

\bibitem{meir1992landauer}
Meir Y and Wingreen N~S 1992 {\em Physical review letters\/} {\bf 68} 2512

\bibitem{ness2010generalization}
Ness H, Dash L and Godby R 2010 {\em Physical Review B\/} {\bf 82} 085426

\end{thebibliography}

\end{document}